\documentclass[preprint,showpacs,preprintnumbers,amsmath,amssymb]{revtex4}
\usepackage{graphicx}
\usepackage{psfig}
\begin{document}
\title{Non-linear magneto-transport
through small quantum dots at strong intra-dot correlations}
\author{I. Sandalov}
\affiliation{Department of Condensed Matter Physics, Royal Institute of Technology,
Electrum 229, SE-164 40 Stockholm-Kista, Sweden}
\affiliation{Max-Planck-Institut f\"ur Physik komplexer
Systeme, D-01187 Dresden, Germany}
\author{R. G. Nazmitdinov}
\affiliation{Departament de F{\'\i}sica,
Universitat de les Illes Balears, E-07122 Palma de Mallorca, Spain}
\affiliation{Bogoliubov Laboratory of Theoretical Physics,
Joint Institute for Nuclear Research, 141980 Dubna, Russia}
\date{\today}
\begin{abstract}
Nonlinear transport through a quantum dot is studied in the limit
of weak and strong intra-dot Coulomb interaction. For the latter
regime the nonequilibrium self-consistent mean field equations for
energies and spectral weights of one-electron transitions are
formulated. The  increase of the bias-voltage window leads to
a strong deviation from Gibbs statistics: the populations of states
involved into a tunnelling are equalizing in this limit even at
low temperature.  For a  symmetric coupling of a quantum dot to two leads
we provide simple analytical relations between heights of the current 
steps and degeneracy of a
spectrum in a two-dimensional parabolic dot in a perpendicular magnetic field
in the both regimes.
\end{abstract}
\maketitle

\section{Introduction}
Nowadays, quantum dots (QDs) are among common experimental tools
to study properties of correlated electron systems at atomic
scale, in a controllable manner. In particular, transport
measurements through QDs provide a rich information on internal
dynamics governed by external gate and bias voltages, and by the
degree of coupling of the dot to the source and drain electrodes.
Depending on the above conditions, a spectrum of a quantum dot,
extracted from the single-electron spectroscopy, manifests a
variety of features from shell structure to chaotic dynamics (see
for review Refs.\onlinecite{kou,RM}).

One of the simplest quantum devices to study transport consists of
two metallic contacts, attached via insulating/vacuum barriers to
a QD. Combination of pieces with very different physical
properties suggests that theoretical descriptions may require
different approximations in accounting for the interactions in the
constituents. In addition, the openings affects an intrinsic
structure  of initially  closed quantum dot. Evidently, the
energies of many-electron transitions (levels) acquire a width and
both single- and many-particle processes contribute to the
life-time of the levels. Moreover, the levels can be shifted due
to the coupling of the QD to the leads. The shift can also be
spin-sensitive due to many-electron nature of the electron
transitions involved \cite{PRL}.

 For small QDs, when the size quantization of the
carrier motion becomes important, and at low enough carrier density
the strong electron correlations (SEC) affect the electron
transport through the dot. At the weak dot-lead coupling and
relatively small temperatures ($\alt 200$ mK), the intra-dot SEC
are dominant in the tunnelling. At this (Coulomb blockade) regime,
the charging energy $E_C$, that needed to add an electron to the
dot with N electrons, considerably exceeds the level spacing and
the thermal energy $k_BT$. According to the classical Coulomb
blockade picture, the transport is only allowed when  the N and
(N+1) states are both energetically accessible (cf
Ref.\onlinecite{tr}). The conductance usually shows a Breit-Wigner
resonance as a function of gate voltage indicating that electrons
pass by a discrete level in the quantum dot by a resonant
tunnelling. At lower temperature the co-tunnelling events
predicted in Ref.\onlinecite{Av} become visible; they involve the
simultaneous tunnelling of two or more electrons over virtual
states, giving rise to the current inside the Coulomb diamond
\cite{Fran}. With a further decrease of the temperature,
Kondo-type effects can develop if the system has quasi-degenerate
localized states. For example, a weak current through QD, observed
within the Coulomb blockade regime \cite{KondoEff}, is interpreted
as a Kondo effect (see for a review Ref.\onlinecite{PG}).

Recently, in experiments with a small quantum dot in the
Coulomb-blockade regime \cite{ihn}, a fine structure was observed
in the conductance as a function of  gate voltage versus
source-drain voltage. It was suggested that this phenomenon is
mainly due to a co-tunnelling \cite{Av}. The theory of
co-tunnelling \cite{Av} neglects, however, specific quantum
effects caused by a strong Coulomb interaction (SCI), which affect
the transition energies and the tunnelling rates. On the other
hand, numerous papers devoted to quantum effects in transport
through QDs are focused on the analysis of Kondo phenomenon in
{\it single-level} QD models (cf Refs.\onlinecite{Kondo}) in a
linear-response regime, which is well understood nowadays.
However,  the single-electron spectroscopy clearly indicates that
a shell structure of small dots plays an essential role in the
transport and weak dot-lead coupling \cite{tr} when the
temperature is still low, but above the Kondo temperature. This
becomes especially evident, when the confining energy exceeds the
charging energy \cite{ota} (hereafter, this regime is called  a
weak Coulomb interaction (WCI) regime). One of our goals is to
present a self-consistent approach for a {\it nonlinear} transport
through a {\it multilevel} quantum dot in the SCI regime. We will
demonstrate that even in this regime a shell structure of a small
dot can be extracted from the analysis of nonlinear current as a
function of a finite bias voltage. We will show also that the
conductance inside of the "Coulomb diamond" is governed by
spectral weights of one-electron transitions, that complement the
co-tunnelling picture.

In this paper we consider the transport through a small QD with a
parabolic effective confining potential. As an input, we use the
intrinsic structure of a {\it closed} dot.  The use of the
parabolic confinement is well justified by experimental
observations, for example, magic numbers of a two-dimensional
harmonic oscillator \cite{tar} and a fulfilment of the Kohn
theorem \cite{Kohn} in far-infrared spectroscopy experiments
\cite{SM}.  The eigen problem of interacting electrons
in a closed parabolic QD can be solved approximately within an
unrestricted Hartree-Fock approximation (cf Ref.\onlinecite{pl03})
or local density approximation to a density functional theory
\cite{RM}. When the confinement dominates  the Coulomb
interaction, the eigen spectrum is well approximated by the
Fock-Darwin levels \cite{fock}. Recent experiments nicely confirm
this fact, indeed (see, for example, experimental results in
Ref.\onlinecite{ota}). This allows to describe a QD within a
single-particle picture of electrons in an effective parabolic
potential (hereafter we denote such states as $|\gamma\rangle$ ).
 We consider the SCI regime in the limit of very strong
intra-dot interaction ("Hubbard" $U\rightarrow\infty$). In this
case  the Hamiltonian of QD at the SCI regime is
diagonalized by the states $\left| 0\right\rangle $ and
$\left|\gamma\right\rangle $, and only the  transitions between empty and single-electron
occupied QD states contribute into transport.
The assumption that the 
  states $\left| 0\right\rangle $ and
$\left|\gamma\right\rangle $  are well separated by a large
energy gap from two- , three-, {\it etc} correlated states 
 can be easily verified, for
example,  within an analytical model for two-electron parabolic
quantum dot \cite{din}. At zero coupling to leads one-electron
energies coincide in both  the WCI and  SCI regimes. It presents a nice
opportunity to study in a transparent way the role of the Coulomb
interaction within this particular charge sector. Although the
bare energies of  QDs coinside n these two regimes, in the SCI limit
 a hopping of an electron into a QD is determined by the
occupation numbers of  the QD states. We employ here the Hubbard
operator technique \cite{ig} which is specially designed for such
cases. The mean field theory in terms of the Hubbard operators is
quite different from the one constructed for the WCI regime. For
example, the consideration of a single spin flip in the infinite
$U$ Hubbard model shows that the mean field theory of this type
gives the same results with both Gutzwiller wave function and the
three-body Faddeev equations in this limit \cite{RuckSchmR}.

As pointed out above, we will analyze the transport through QD in
non-linear with respect to  the bias voltage regime at the weak
dot-lead coupling. In other words, the transparency of junctions
is assumed to be small. The latter requirement has a pragmatic
basis as well: enormous efforts for a fabrication of QDs are
motivated by the desire to understand the inner structure and to
exploit it in various applications.

Thus, the coupling to the leads will be treated perturbatively.
Below we focus on the transport in the resonant-tunnelling regime,
where the Kondo physics is not involved. In the language of the
Anderson impurity theory, we will be interested in the
intermediate-valence regime,  when  QD levels are located
within the range of the resonant tunnelling. We recall that in
this region such a widely used method as a mean-field
approximation in slave-boson theory \cite{geor} is not applicable,
in contrast to our approach.

The paper is organized as follows:  in Section II we will discuss
our model Hamiltonian, the nonlinear current in terms of  the Hubbard
operators and formulate the mean field approximation. Section III
is devoted  to the derivation of  equations for population numbers and
energy shifts in the diagonal approximation.  The results for the
transport through a parabolic quantum dot in a perpendicular
magnetic field at weak (WCI) and strong (SCI) intra-dot Coulomb
interaction are given in Section IV. The conclusions are finally
drawn in Section V. Two Appendices contain some technical details.
 Some preliminary results of our analysis are presented in
Ref.\onlinecite{JPhysC}

\section{Formalism }
\subsection{Model Hamiltonian and current}
As discussed in the Introduction, we consider the system  which is decribed by
the following Hamiltonian
\begin{equation}
 H=H_{l}+H_{r}+H_{QD}+H_{t},
\label{H1}
\end{equation}
where
\begin{eqnarray}
&&H_{\lambda} =\sum_{k\sigma \in \lambda}
\varepsilon_{\lambda k\sigma}\,c_{\lambda k\sigma}^{\dagger}
c_{\lambda k\sigma}\nonumber\\
&&H_{QD}=\sum_{\gamma}\varepsilon_{\gamma}n_\gamma+
\sum_{\gamma\neq \gamma^{\prime}} U_{\gamma \gamma^{\prime}}n_\gamma n_{\gamma^{\prime}},
\quad n_\gamma=d_\gamma^{\dagger}d_\gamma \nonumber\\
&&H_{t} =\sum_{k\sigma\in \lambda,\lambda,\gamma}
\left(v_{k\sigma,0\gamma}^{\lambda}\,c_{\lambda k\sigma}^{\dagger}d_\gamma+h.c.\right)\nonumber
\end{eqnarray}
The term $H_{\lambda}$ describes noninteracting electrons with the
energy $\varepsilon_{\lambda k\sigma}$, wave number $k$ and spin
$\sigma$ in the lead $\lambda=l,r$. The closed dot is modelled by
$H_{QD}$. Here  $\varepsilon_{\gamma}$ is the energy of electron
with a set of single-electron quantum numbers $\gamma$ in the
confining potential of the closed QD at $U=0$.
 Tunnelling between the dot and leads is described
by the term $H_{t}$; the matrix elements
$v_{k\sigma,0\gamma}^{\lambda}$ here couple  the left and the right
contacts with  a QD. The ratio of level spacing of noninteracting
electrons in a QD to "Hubbard $U$",
$\left|\varepsilon_{\gamma}-\varepsilon_{\gamma^{\prime}}\right|/U=
\delta\varepsilon_{\gamma \gamma^{\prime}}/U$ determines whether
 the intra-dot electrons are in the regime of weak
($\delta\varepsilon_{\gamma \gamma^{\prime}}/U >1$) or strong
($\delta\varepsilon_{\gamma \gamma^{\prime}}/U <1$) Coulomb
interaction.

The Hamiltonian of QD is transformed into a diagonal form by
introducing the Hubbard operators \cite{hub} (see also Appendix
\ref{app1}). As a result, the Hamiltonian (\ref{H1}) acquires the
form:

\begin{eqnarray}
& H = \sum_{k\sigma\lambda}\varepsilon_{\lambda k\sigma}
c_{\lambda k\sigma}^{\dagger}c_{\lambda
k\sigma}+\varepsilon_0Z^{00}+
\sum_{\gamma}\varepsilon_{\gamma}Z^{\gamma \gamma}\nonumber\\
& + \sum_{k\sigma \lambda,\gamma} \left(
v_{k\sigma,\gamma}^{\lambda}c_{\lambda k\sigma}^{\dagger
}X^{0\gamma}+h.c.\right) \label{fp7}
\end{eqnarray}

Following Ref.\onlinecite{MW}, we obtain (see Appendix \ref{app2})
for the "left" steady current
\begin{eqnarray}
J_{l}  &=&\frac{ie}{\hbar}\sum_{\gamma \gamma^{\prime}}\int d\omega \Bigg{\{}
\left[\Gamma_{\gamma^{\prime}\gamma}^{l}(\omega)-
\Gamma_{\gamma^{\prime}\gamma}^{r}(\omega)\right]G_{\gamma,\gamma^{\prime}}^{<}(\omega) \nonumber\\
&+&\left[  \Gamma_{\gamma^{\prime}\gamma}^{l}(\omega)
f_{l}(\omega)-\Gamma_{\gamma^{\prime}\gamma}^{r}(\omega)f_{r}(\omega)\right]\times\nonumber\\
&&\times\left[G_{\gamma,\gamma^{\prime}}^{R}(\omega)-G_{\gamma, \gamma^{\prime}}^{A}(\omega)\right]
\Bigg{\}}
\label{curr_in_Hubb}%
\end{eqnarray}

The current (\ref{curr_in_Hubb}) is expressed in terms of Green
functions (GFs) defined for the Hubbard operators :
\begin{eqnarray}
&&G_{\gamma, \gamma^{\prime}}^{\eta}(\omega)=\int d(t-t^{\prime})
e^{i\omega(t-t^{\prime})}G_{\gamma, \gamma^{\prime}}^{\eta}(t,t^{\prime}),\;\eta
=<,>,R,A \nonumber\\
&&G_{\gamma, \gamma^{\prime}}^{<}(t,t^{\prime})  \equiv
i\left\langle X^{\gamma^{\prime}0}(t^{\prime})X^{0\gamma}(t)\right\rangle \nonumber\\
&&G_{\gamma, \gamma^{\prime}}^{>}(t,t^{\prime})\equiv -i\left\langle X^{0\gamma}(t)
X^{\gamma^{\prime}0}(t^{\prime})\right\rangle\nonumber\\
&&G_{\gamma, \gamma^{\prime}}^{R/A}(t,t^{\prime}) \equiv
\mp i\theta(\pm t\mp t^{\prime})\left\langle \left\{X^{0\gamma}(t),X^{\gamma^{\prime}0}
(t^{\prime})\right\}  \right\rangle
\label{dfGFs}
\end{eqnarray}
The  GFs depend on difference of times (or one  frequency $\omega
$) because we consider the steady states only.  Note that for 
a wide conduction band, i.e., when the bandwidth is
much larger than any other parameter in the problem, the width
function

\begin{equation}
\Gamma_{\gamma\gamma^{\prime}}^{\lambda}(\omega)=\pi\sum_{\,k\sigma}
v_{\gamma^{\prime},k\sigma}^{\lambda}\delta(\omega-
\varepsilon_{\lambda k\sigma})v_{k\sigma,\gamma}^{\lambda}
\label{width}
\end{equation}
weakly depends on $\omega$ and will be replaced by a constant when
we will consider a particular model (see also Appendices
\ref{app1},\ref{app2})

Thus, we have to find the GFs $G_{\gamma,
\gamma^{\prime}}^{\eta}$. As well-known, the Dyson equation does
not exist for any many-electron GF (particularly, for the GF
defined on the Hubbard operators), since the (anti)commutation of
many-electron operators (see Eq.(\ref{Commut})) generates not a
"c"-number as in case of fermions or bosons, but an operator
again. The  perturbation theory for the many-electron GFs reflects
this fact by generating the additional graphs compared to the
conventional theory for fermions or bosons (examples are given in
Ref.\onlinecite {ig}; also see a Sec.IIb). Therefore, those of
results of Refs.\onlinecite{MW,JWM}, which are derived on the
basis of a Dyson equation for the retarded and advanced GFs,
should be re-inspected. As will be seen below,  the formulas
analogous to the ones derived in Refs.\onlinecite{MW,JWM} are
valid only within the \ well-known Hubbard-I approximation and
mean-field approximations (MFA) \cite{ig} (Ref.\onlinecite{ig} will
be referred below as I).  The same approximation is  sufficient 
for calculation of the transport properties at small transparencies 
of the junctions too.

 Let us introduce the full width $\Gamma $  and dimensionless partial widths $\alpha $:
\begin{equation}
\Gamma_{\gamma^{\prime}\gamma}=\Gamma_{\gamma^{\prime}\gamma}^{l}+
\Gamma_{\gamma^{\prime}\gamma}^{r}, \quad
\alpha_{\gamma^{\prime}\gamma}^{\lambda}=
\Gamma_{\gamma^{\prime}\gamma}^{\lambda}/\Gamma_{\gamma^{\prime}\gamma},
\label{alpha}%
\end{equation}
When the coupling to leads is so small that the widths $\Gamma$ of
the  transitions are much smaller than the temperature $T$,
$\beta \Gamma \ll 1$ (where $ \beta = 1/T$) and the level spacing,
$\Gamma/\delta \epsilon_{\gamma \gamma^{\prime}}\ll1$, the mixing
of levels due to non-diagonal terms does not influence the physics
of transport. Namely this range of parameters is exploited usually
in experiments and will be in the scope of our interests (starting
from Sec.III). Neglecting the non-diagonal terms leads to the {\it
diagonal approximation}, where
$\Gamma_{\gamma^{\prime}\gamma}=\delta_{\gamma^{\prime}\gamma}\Gamma_{\gamma}$,
$\alpha_{\gamma
\gamma^{\prime}}^{\lambda}=\delta_{\gamma^{\prime}\gamma}\alpha_{\gamma}^{\lambda}$.
The expression for the current becomes:
\begin{eqnarray}
J & = &\frac{ie}{\hbar}\sum_{\gamma}\Gamma_{\gamma \gamma}\int
d\omega \Bigg{\{}
\left[\alpha_{\gamma}^{l}-\alpha_{\gamma}^{r}\right]  G_{\gamma,\gamma}^{<}(\omega)\nonumber\\
& +&\left[  \alpha_{\gamma}^{l}f_{l}(\omega)-\alpha_{\gamma}^{r}f_{r}%
(\omega)\right]  \left[  G_{\gamma,\gamma}^{R}(\omega)-G_{\gamma,\gamma}^{A}(\omega)\right]
\Bigg {\}}
\label{diagCurr}
\end{eqnarray}
The advantage of the Wingreen and Meir formulation \cite{MW} lies
in the fact that   the current (in our case
Eq.(\ref{curr_in_Hubb})) is expressed in terms of the GFs of a QD
only. We recall that this result can be obtained only for
non-interacting conduction electrons. Thus, our next-step goal is
to derive equations for GFs $G_{\gamma,p^{\prime}0}^{<,R,A}$ in
the MFA.

\subsection{Exact equations on imaginary-time axis}
\label{sec-gf} 

Our derivation is based on the approach
\cite{PRL,ig}  that employs  the ideas suggested by Schwinger
\cite{schwinger} and developed further by Kadanoff and Baym
\cite{KB}. Since we consider only the intra-dot Coulomb
interaction, the conduction-electron subsystems are linear and can
be integrated \ out.  Instead an additional
effective interactions between electrons in a QD arises. Following
Ref.\onlinecite{KB}, equations for the QD subsystem will be
written in terms of functional derivatives with respect to
external auxiliary sources. The derivation of the equations will
be performed in three steps. Firstly, we derive exact equations
for the QD GFs on the imaginary-time axis. Secondly, the MFA will
be formulated. Thirdly, we will perform a continuation of these
equations onto the real-time axis.

The time evolution of the Hubbard operator $X^{0\gamma}$ is defined by the
commutator
\begin{eqnarray}
[X^{0\gamma},H] & = &\Delta_{\gamma 0}^{(0)}X^{0\gamma}+\\
&+&\sum_{k\sigma\in \lambda,\lambda,\gamma_{1}\neq \gamma}\,v_{\gamma_{1},k\sigma}^{\lambda}
\left(\delta_{\gamma_{1}\gamma}Z^{00}+Z^{\gamma_{1}\gamma}\right)  c_{\lambda k\sigma}\nonumber,
\label{gf4}
\end{eqnarray}
which contains non-linear terms. These terms produce GFs of the
type $-i\langle T\left( \delta_{\gamma^{\prime}\gamma}Z^{00}(t)+
Z^{\gamma^{\prime}\gamma}(t)\right)c_{\lambda
k^{\prime\prime}\sigma}(t) c_{\lambda
k^{\prime}\sigma}^{\dagger}(t^{\prime})\rangle$. Below we omit the
summation signs: summation over repeating indices is implied
unless otherwise  stated.

To proceed further, we introduce the auxiliary sources $U^{00}%
(t),U^{\gamma \gamma^{\prime}}(t)$ into a definition of the GFs:
\begin{eqnarray}
&&G_{AB}(t,t^{\prime})=
\frac{1}{i}\left\langle TA(t)B(t^{\prime})\right\rangle _{U}
=\frac{1}{i}\frac{\left\langle TA(t)B(t^{\prime})S(t_{0},\beta)\right\rangle }
{\left\langle TS(t_{0},\beta)\right\rangle},\nonumber\\
&&S(t_{0},\beta) \equiv T\exp \Bigg{\{}-i\int_{t_{0}}^{t_{0}-i\beta}
dt_{1}\Bigg{[}  U^{00}(t_{1})Z^{00}(t_{1})+\nonumber\\
&&+\sum_{\gamma \gamma^{\prime}}U^{\gamma \gamma^{\prime}}(t_{1})
Z^{\gamma \gamma^{\prime}}(t_{1})
\Bigg{]}\Bigg{\}}.
\label{dfG}
\end{eqnarray}
Here $A,B=c_{\lambda k\sigma},X^{0\gamma}$; ($\lambda=l,$) for for
the left and  ($\lambda=r$) for the right contacts.  We will use
the following notations:
\begin{widetext}
\begin{equation}
\left(
\begin{array}
[c]{cc}
C_{k\sigma,k^{\prime}\sigma^{\prime}}(t,t^{\prime}) &
G_{k\sigma,\gamma^{\prime}}(t,t^{\prime})\\
G_{\gamma,k^{\prime}\sigma^{\prime}}(t,t^{\prime}) & G_{\gamma,\gamma^{\prime}}(t,t^{\prime})
\end{array}
\right)  =\frac{1}{i}\left(
\begin{array}
[c]{cc}
\left\langle Tc_{\lambda k\sigma}(t)c_{\lambda^{\prime} k^{\prime}\sigma^{\prime}}^{\dagger}
(t^{\prime})\right\rangle _{U} & \left\langle Tc_{\lambda k\sigma}(t)X^{\gamma^{\prime}0}
(t^{\prime})\right\rangle _{U}\\
\left\langle TX^{0\gamma}(t)c_{\lambda^{\prime}k^{\prime}\sigma^{\prime}}^{\dagger
}(t^{\prime})\right\rangle _{U} & \left\langle TX^{0\gamma}(t)X^{\gamma^{\prime}%
0}(t^{\prime})\right\rangle _{U}%
\end{array}
\right).
\label{MatrG}
\end{equation}
\end{widetext}
 The equation of motion for
the GFs $C_{k\sigma,k^{\prime}\sigma^{\prime}}(t,t^{\prime})$ and
$G_{k\sigma,\gamma^{\prime}}(t,t^{\prime})$ are simple:
\begin{eqnarray}
&&\left(i\partial_{t}-\varepsilon_{\lambda k\sigma}\right)
C_{k\sigma,k^{\prime}\sigma^{\prime}}(t,t^{\prime})
= \delta_{kk^{\prime}}\delta_{\sigma\sigma^{\prime}}+\nonumber\\
&&+\sum_{\gamma^{\prime\prime}}\,v_{k\sigma,\gamma^{\prime\prime}}^\lambda
G_{\gamma^{\prime\prime},k^{\prime}\sigma^{\prime}}(t,t^{\prime})
\label{C}\\
&&\left(i\partial_{t}-\varepsilon_{\lambda k\sigma}\right)
G_{k\sigma,\gamma^{\prime}}(t,t^{\prime}) = \sum_{\gamma^{\prime\prime}}\,
v_{k\sigma,\gamma^{\prime\prime}}^\lambda G_{\gamma^{\prime\prime},\gamma^{\prime}}(t,t^{\prime})\nonumber\\
\label{G_cX}%
\end{eqnarray}
As seen from Eq.(\ref{C}), a zero GF $C_{k\sigma,k^{\prime}\sigma^{\prime}}^{(0)}$
satisfies the equation
\begin{eqnarray}
\int dt_{1}\sum_{k^{\prime\prime}}\{[i\partial_{t}-
\varepsilon_{\lambda k\sigma}]\delta(t-t_{1})\}C_{k\sigma,k^{\prime}\sigma^{\prime}}^{(0)}
(t,t^{\prime})&\equiv&\\
\delta_{kk^{\prime}}\delta_{\sigma
\sigma^{\prime}}\delta(t-t^{\prime})\nonumber
\label{zeroC}%
\end{eqnarray}
Hereafter the integrals are taken between $t_{0}$ and
$t_{0}-i\beta$, if the other limits are not specified explicitly;
here $t_{0}$ is an arbitrary moment. With the aid of this
equation, Eqs.(\ref{C}),(\ref{G_cX}) can be cast in the form of
integral equations:
\begin{eqnarray}
C_{k\sigma,k^{\prime}\sigma^{\prime}}(t,t^{\prime})
& = &C_{k\sigma}^{(0)}(t,t^{\prime})\delta_{kk^{\prime}}
\delta_{\sigma \sigma^{\prime}}+\\
&+&\sum_{\gamma^{\prime\prime}}\int dt_{1}
C_{k\sigma}^{(0)}(t,t_{1}) v_{k\sigma,\gamma^{\prime\prime}}^{\lambda}
G_{\gamma^{\prime\prime},k^{\prime}\sigma^{\prime}}(t_{1},t^{\prime})\nonumber
\label{Cint}\\
G_{k\sigma,\gamma^{\prime}}(t,t^{\prime})& = & \sum_{\gamma_{2}}
\,\int dt_{1}C_{k\sigma}^{(0)}(t,t_{1})v_{k\sigma,\gamma_{2}}^{\lambda}
G_{\gamma_{2},\gamma^{\prime}}(t_{1},t^{\prime})\nonumber\\
\label{G_cXint}
\end{eqnarray}
Analogously, using the equation for the GF
$G_{\gamma^{\prime\prime},k^{\prime}\sigma^{\prime}}(t_{1},t^{\prime})$ with respect
to the right time $t^{\prime}$, we obtain
\begin{equation}
G_{\gamma^{\prime\prime},k^{\prime}\sigma^{\prime}}(t,t^{\prime})=
\sum_{\gamma_{1}}\,\int dt_{1}G_{\gamma^{\prime\prime},\gamma_{1}}(t,t_{1})
v_{\gamma_{1},k^{\prime}\sigma^{\prime}}^\lambda C_{k^{\prime}\sigma^{\prime}}^{(0)}
(t_{1},t^{\prime})
\label{G_Xc_int}
\end{equation}
The equation for $G_{\gamma, \gamma^{\prime}}(t,t^{\prime})=-i\left\langle
TX^{0\gamma}(t)X^{\gamma^{\prime}0}(t^{\prime})\right\rangle _{U}$ has, however, more
complex form:
\begin{eqnarray}
&&  \left[ \delta_{\gamma^{\prime}\gamma}i\partial_{t}-\Delta_{\gamma \gamma_{1}}^{(0)}(t)\right]
G_{\gamma_{1},\gamma^{\prime}}(t,t^{\prime})=\delta(t-t^{\prime})P^{\gamma \gamma^{\prime}}(t)+\nonumber\\
&&\sum_{k\sigma \in \lambda,\lambda,\gamma_{1}}\,v_{\gamma_{1},k\sigma}^{\lambda}
\frac{1}{i}\left\langle T\left(\delta_{\gamma_{1}\gamma}Z^{00}+Z^{\gamma_{1}\gamma}\right)
c_{\lambda k\sigma}(t)X^{\gamma^{\prime}0}(t^{\prime})\right\rangle _{U}, \nonumber\\
\label{EqForG_XB}
\end{eqnarray}
where
\begin{eqnarray}
&&P^{\gamma \gamma^{\prime}}(t)\equiv\left\langle T\left\{ X^{0\gamma}(t),X^{\gamma^{\prime}0}(t)\right\}
\right\rangle _{U}\\
&& \Delta_{\gamma \gamma_{1}}^{(0)}(t)= \Delta_{\gamma}^{(0)}\delta_{\gamma_{1}\gamma}+
\left[U^{\gamma \gamma_{1}}(t)-\delta_{\gamma_{1}\gamma}U^{00}(t)\right]\nonumber
\label{Notat_P_Delt}
\end{eqnarray}
Here $\Delta_{\gamma}^{(0)}\equiv\varepsilon_\gamma
-\varepsilon_0$  is the transition energy between  the
single-electron state and the one without electrons in a QD. We
choose $\varepsilon_0 =0$. The term $\left( U^{\gamma
\gamma^{\prime}}(t)-\delta_{\gamma^{\prime}\gamma}U^{00}(t)\right)$
is due to the differentiation of $S(t_{0},\beta)$ (cf
Ref.\onlinecite{KB}).

Below, we will proceed under the standard assumption that
the interaction is absent at infinitely remote time $t_{0}\rightarrow-\infty$
and is switched on adiabatically. This assumption does not affect
our approach, since we study a stationary regime.
In this case we use the relation:
\begin{eqnarray}
&\langle TZ^{\xi}(t)X^{0\gamma}(t)B(t^{\prime})\rangle_{U}=
\nonumber\\
&\left(  \langle TZ^{\xi}(t)\rangle+i\frac{\delta}{\delta U^{\xi}(t)}\right)\langle
TX^{0\gamma}(t)B(t^{\prime})\rangle_{U}
\label{KBtrick}
\end{eqnarray}
that follows from a functional derivative $\delta/\delta U^{\xi}(t)$ of
Eq.(\ref{dfG}) (here $\xi=[00],[\gamma \gamma^{\prime}]$).

 With the aid of Eq.(\ref{KBtrick}), using Eq.(\ref{G_cXint}) and the definition
for the effective interaction via conduction electrons
\begin{equation}
V_{\gamma_1 \gamma_2}(t,t_1)=\sum_{k\sigma \in \lambda,\lambda}
v_{\gamma_1,k\sigma}^{\lambda}C_{k\sigma}^{(0)}(t,t_1)
v_{k\sigma,\gamma_2}^{\lambda}
\label{effint}
\end{equation}
 (see also Eq.(\ref{effin})), we transform Eq.(\ref{EqForG_XB}) to the form
\begin{eqnarray}
&&\left[\delta_{\gamma \gamma_{1}}i\partial_{t}-\Delta_{\gamma \gamma_{1}}^{(0)}(t)\right]
G_{\gamma_{1},\gamma^{\prime}}(t,t^{\prime})=\delta(t-t^{\prime})P^{\gamma \gamma^{\prime}}(t)\nonumber\\
&&  +\sum_{\gamma_{2}\gamma_{1}}\,\int dt_{1}
V_{\gamma_{1}\gamma_{2}}(t,t_{1})\Bigg{(}P^{\gamma \gamma_{1}}(t)+
i\delta_{\gamma \gamma_{1}}\frac{\delta}{\delta U^{00}(t)}+\nonumber\\
&& + i\frac{\delta}{\delta U^{\gamma_{1}\gamma}(t)}\Bigg{)}
G_{\gamma_{2},\gamma^{\prime}}(t_{1},t^{\prime})
\label{EqForG_XX}
\end{eqnarray}
Here, we used also that $\delta C_{k\sigma}^{(0)}/\delta U=0$.

Eq.(\ref{EqForG_XX}) is the exact equation for the GF
$G_{\gamma,\gamma^{\prime}}(t,t^{\prime})$
on the imaginary time axis. Since all other GFs are expressed in terms of the
$G_{\gamma^{\prime\prime},\gamma}$, the iteration of Eq.(\ref{EqForG_XX}) with respect
to the effective interaction generates a full perturbation theory.
However, within this formulation of the theory, the continuation from
the imaginary time axis to the real one should be performed in each term of
the expansion.

The well-known  ''Hubbard-I'' (HI) approximation can be obtained
from  Eq.(\ref{EqForG_XX}) by putting therein $\delta
G_{\gamma^{\prime\prime},\gamma}/\delta U=0$.  Let us  define a
zero \emph{locator} $D_{\gamma,
\gamma^{\prime}}^{(0)}(t,t^{\prime})$
\begin{eqnarray}
&&\int dt_{1} \left\{\left[\delta_{\gamma \gamma_{1}}i\partial_{t}-
\Delta_{\gamma \gamma_{1}}^{(0)}(t)\right] \delta(t-t_{1})\right\}
D_{\gamma_{1},\gamma_{2}}^{(0)}(t_{1},t^{\prime})\nonumber\\
&&  \equiv\int
dt_{1}\left[D^{(0)-1}(t,t_1)\right]_{\gamma,\gamma_{1}}
D_{\gamma_{1},\gamma_{2}}^{(0)}(t_{1},t^{\prime})\nonumber\\
&&=\delta(t-t^{\prime})\delta_{\gamma \gamma_{2}}
\label{dfD0}
\end{eqnarray}
In  the HI approximation, the locator is defined by a solution to the equation:
\begin{eqnarray}
&&\int dt_{1}\sum_{\gamma_{1}}\Bigg{\{}\left[\delta_{\gamma \gamma_{1}}i\partial_{t}
-\Delta_{\gamma \gamma_{1}}^{(0)}(t)\right]\delta(t-t_{1})-\nonumber\\
&&S_{\gamma,\gamma_{1}}^{HI}(t,t_{1})\Bigg{\}} D_{\gamma_{1},\gamma_{3}}^{HI}(t_{1},t^{\prime})
=\delta_{\gamma \gamma_{3}}\delta(t-t^{\prime}),
\label{D_HIA}
\end{eqnarray}
where
\begin{equation}
S_{\gamma,\gamma_{1}}^{HI}(t,t_{1})=\sum_{\gamma_{2}}
P^{\gamma \gamma_{2}}(t)\,V_{\gamma_{2}\gamma_{1}}(t,t_{1})
\label{S_HIA}
\end{equation}
Using the above equations, we rewrite  Eq.(\ref{EqForG_XX}) in the form
\begin{eqnarray}
&&\int dt_{1}\sum_{\gamma_{1}}\Bigg{\{}\left[D^{HI,-1}(t_{1},t^{\prime})\right]_{\gamma,\gamma_{1}}-\\
&&-\tilde{S}_{\gamma,\gamma_{1}}(t,t_{1})\Bigg{\}} G_{\gamma_{1},\gamma^{\prime}}(t_{1},t^{\prime})
=\delta(t-t^{\prime})P^{\gamma \gamma^{\prime}}(t^{\prime})\nonumber
\label{EqForG}
\end{eqnarray}
with
\begin{eqnarray}
& \int dt_{1}\sum_{\gamma_{1}}\tilde{S}_{\gamma,\gamma_{1}}(t,t_{1})
G_{\gamma_{1},\gamma^{\prime}}(t_{1},t^{\prime})\equiv
\label{StildeG}\\
&i\sum_{\gamma_{2}\gamma_{1}}\,\int
dt_{1}V_{\gamma_{1}\gamma_{2}}(t,t_{1}) \left( \delta_{\gamma
\gamma_{1}}\frac{\delta G_{\gamma_{2},\gamma^{\prime}}
(t_{1},t^{\prime})}{\delta U^{00}(t)} +\frac{\delta
G_{\gamma_{2},\gamma^{\prime}}(t_{1},t^{\prime})}{\delta
U^{\gamma_{1}\gamma}(t)}\right )\nonumber
\end{eqnarray}
One may notice that Eq.(\ref{StildeG}) has a simple form in symbolic notations
\begin{equation}
D_{HI}^{-1}G+V\frac{\delta G}{\delta U}=P
\label{shortEqG}
\end{equation}
This equation suggests that the \emph{full locator} $D$ can be defined
by the equation
\begin{equation}
\sum_{\gamma_{1}}\int dt_{1}[D(t,t_{1})]_{\gamma,\gamma_{1}}^{-1}G_{\gamma_{1},\gamma^{\prime}}
(t_{1},t^{\prime})=\delta(t-t^{\prime})P^{\gamma \gamma^{\prime}}(t^{\prime}),
\label{mft1}
\end{equation}
where
\[
\lbrack D(t,t_{1})]_{\gamma,\gamma_{1}}^{-1}\equiv\lbrack D^{0}(t,t_{1})]_{\gamma,\gamma_{1}}^{-1}
-S_{\gamma,\gamma_{1}}(t,t_{1})]
\]
and the full self-operator is defined as
\begin{equation}
S_{\gamma,\gamma_{1}}=S_{\gamma,\gamma_{1}}^{HI}+\tilde{S}_{\gamma,\gamma_{1}}
\label{fullS}
\end{equation}
The magnitude $S_{\gamma,\gamma_{1}}$ has been named in I as ''a self-operator'' in
order to  distinguish it from the standard self-energy operator. In fact,
the magnitude $\tilde{S}_{\gamma,\gamma_{1}}(t,t^{\prime})\,$ renormalizes
the transition energies in a QD as well as the end-factors
$P^{\gamma \gamma^{\prime}}(t)$. Indeed, one can see from Eq.(\ref{mft1}) that
\begin{equation}
G_{\gamma_{1},\gamma^{\prime}}(t,t^{\prime})=D_{\gamma_{1},\gamma_{2}}(t,t^{\prime})
P^{\gamma_{2}\gamma^{\prime}}(t^{\prime})
\label{G=DP}%
\end{equation}
and, therefore, $P^{\gamma_{2}\gamma^{\prime}}(t^{\prime})$ enters into
the equations for the GFs \emph{under} the sign of the functional derivative.
Being an expectation value from the anticommutator of \emph{many-electron} operators,
it is not a constant (like in the cases of Bose- or Fermi-operators).
Therefore, its functional derivative is nonzero and generates a sub-set of
graphs that do not appear in standard techniques for fermions and bosons.
These graphs describe\emph{\ kinematic} interactions.

\subsection{The energy shifts (imaginary time).}

An effective interaction of the  intra-dot states via conduction
electrons arises in both  the WCI and SCI regimes. In  both cases the
lowest orders of perturbation theory contains the terms with
integration over wide energy region of the continuous spectrum
and, therefore,  it results in finite widths of
the intra-dot states.  Particularly, in  the SCI regime the width is
provided already by the Hubbard-I approximation.  Below we will show that
in contrast to the WCI regime, in the SCI case the
width depends on the population of the many-electron states which
are involved into single-electron transition in question. This
fact is more or less  trivial consequence of non-Fermi/Bose
commutation relations between Hubbard operators.  A non-trivial
consequence of it is kinematic interaction that shift
transitions energies $\Delta_{\gamma}^{(0)}$ in a QD. The
correlation-caused shift arises in  a natural way within 
 the mean-field approximation in slave-boson theory (see, e.g.,
Ref.\cite{langreth}).
This approach misses, however, the spin and orbital sensitivity of
the shifts for different levels. This obstacle can be resolved by
means of the Hubbard operators formalism (cf I and
Ref.\onlinecite{PRL}). In the present paper we use also the  mean
field approximation, which is similar in spirit to the one
of Ref.\onlinecite{langreth}. The details and physical meaning of
our MFA, however, are different.

The MFA for the Hubbard operator GFs have been defined in I on
 the imaginary-time axis by two conditions: first, the dynamical
scattering on Bose-like excitations is neglected, {\it i.e.},
$\delta P/\delta U=0$ and, second, full vertex $\Gamma\equiv\delta
D^{-1}/\delta U\ $ is replaced by zero one, $\Gamma_{0}\equiv\delta
D_{0}^{-1}/\delta U$. In other words,  the MFA omits the fluctuations
around the stationary solutions. The advantage of formulation of
the theory  in terms of Hubbard operators  compared the
slave-boson technique  consists in the following. The MFA can be
derived for a general case of an arbitrary Hubbard-operator
algebra, or, in physical terms, for the case of arbitrary number
and a type of (Bose- or Fermi-like) transitions (see  I). 
 Besides, one can go beyond the MFA in a systematic way. 
For example, in  the presence of the long-range part of  the Coulomb 
interaction one can reformulate  theory in terms of screened interaction (see
Ref.\cite{condmatII} ).

In order to develop the MFA we have to extract from the self-operator
$\tilde{S}$ the local-in-time contribution in the lowest order with respect to
the effective interaction $V$. The self-operator $\tilde{S}$
(see Eq.(\ref{StildeG})) is already proportional to the interaction
$V_{\gamma \gamma^{\prime}}$.
Therefore, one should calculate only the derivative
$\delta G/\delta U$ in the lowest order with respect to the interaction.
We recall that the interaction $V_{\gamma^{\prime}\gamma}(t,t_{1})$
does not depend on $U^{\xi}$. Thus, the derivative is
\begin{eqnarray}
& \frac{\delta G_{\gamma_{2},\gamma^{\prime}}(t_{1},t_{2})}{\delta U^{\xi}(t)}%
=\frac{\delta\left[  D_{\gamma_{2},\gamma_{5}}(t_{1},t_{2})P^{\gamma_{5}\gamma^{\prime}}%
(t_{2})\right]  }{\delta U^{\xi}(t)}\label{derivG}\\
&  =\frac{\delta D_{\gamma_{2},\gamma_{5}}(t_{1},t_{2})}{\delta U^{\xi}(t)}%
P^{\gamma_{5}\gamma^{\prime}}(t_{2})+D_{\gamma_{2},\gamma_{5}}(t_{1},t_{2})\frac{\delta
P^{\gamma_{5}\gamma^{\prime}}(t_{2})}{\delta U^{\xi}(t)}
\nonumber
\end{eqnarray}
 The second term in Eq.(\ref{derivG}) expresses the dynamical re-scattering process
\begin{eqnarray*}
&  \delta P^{\gamma_{5},\gamma^{\prime}}/\delta U^{\xi}
\propto\left\langle TZ^{00}Z^{\xi}\right\rangle -\\
&-\left\langle TZ^{00}\right\rangle \left\langle TZ^{\xi}\right\rangle +
\left\langle TZ^{\gamma_{4}\gamma_{5}}Z^{\xi}\right\rangle -
\left\langle TZ^{\gamma_{4}\gamma_{5}}
\right\rangle \left\langle TZ^{\xi}\right\rangle ,
\end{eqnarray*}
where the Fermi-like excitation, described by the locator $D$,
scatters on the Bose-like excitation, described by the GFs $\delta
P/\delta U$. The diagonal correlators
$\left\langle TZ^{\gamma \gamma}(t)Z^{\gamma \gamma}(t^{\prime})\right\rangle
-\left\langle TZ^{\gamma \gamma}(t)\right\rangle
\left\langle TZ^{\gamma \gamma}(t^{\prime})\right\rangle$ and
$\left\langle TZ^{00}(t)Z^{00}(t^{\prime})\right\rangle
-\left\langle TZ^{00}(t)\right\rangle \left\langle Z^{00}(t^{\prime})\right\rangle$
describe the fluctuations of the population numbers
  and $\left\langle TZ^{\gamma \gamma}\right\rangle \equiv N_{\gamma}$
and $\left\langle TZ^{00}\right\rangle \equiv N_{0}$, respectively. The
non-diagonal GFs
$\left\langle TZ^{\gamma \gamma^{\prime}}(t)Z^{\gamma^{\prime}\gamma}(t^{\prime})\right\rangle
$ describe the transitions between one-electron states $\gamma$
and $\gamma^{\prime}$. We consider the resonant tunnelling
(non-Kondo regime !)  at low temperature and at small
transparencies of junctions. There the fluctuations of the
population numbers produced by the dynamics can be ignored in this
region of parameters. This makes the MFA applicable.

Taking into account only the first term in Eq.(\ref{derivG}), we obtain

\begin{eqnarray}
&&\tilde{S}_{\gamma,\gamma_{3}}(t,t_{3})\equiv i\sum_{\gamma_{1}\gamma_{2}}\,\int dt_{1}
\int dt_{2}V_{\gamma_{1}\gamma_{2}}(t,t_{1})\times
\nonumber\\
&&\times\Bigg{(}  \delta_{\gamma \gamma_{1}} \frac{\delta
G_{\gamma_{2},\gamma_{4}}(t_{1},t_{2})}{\delta U^{00}(t)}+
\frac{\delta G_{\gamma_{2},\gamma_{4}}(t_{1},t_{2})}{\delta
U^{\gamma_{1}\gamma}(t)} \Bigg{)}
G_{\gamma_{4}\gamma_{3}}^{-1}(t_{2},t_{3})
\nonumber\\
&&=i\sum_{\gamma_{1}\gamma_{2}}\,\int dt_{1}\int
dt_{2}V_{\gamma_{1}\gamma_{2}}(t,t_{1}) \Bigg{(}  \delta_{\gamma
\gamma_{1}} \frac{\delta
D_{\gamma_{2},\gamma_{5}}(t_{1},t_{2})}{\delta U^{00}(t)}+
\nonumber\\
&&+\frac{\delta D_{\gamma_{2},\gamma_{5}}(t_{1},t_{2})}{\delta U^{\gamma_{1}\gamma}(t)}
\Bigg{)} P^{\gamma_{5}\gamma_{4}}(t_{2}) G_{\gamma_{4}\gamma_{3}}^{-1}(t_{2},t_{3})
\label{S1}
\end{eqnarray}
The derivative of the locator $\delta D/\delta U$ can be calculated
using the trick
\begin{equation}
\delta D/\delta U=-D\left(  \delta D^{-1}/\delta U\right)  D,
\label{trick}
\end{equation}
that follows from the equation $\delta\left(  D^{-1}D\right)  /\delta U=0$.
The validity of the trick is well justified, since we consider only stationary processes
and within the assumption that the interaction, which is
absent at remote time $t_{0}\rightarrow-\infty$, is switched on adiabatically.
By means of Eqs.(\ref{G=DP}),(\ref{trick}), we have
\begin{eqnarray*}
&\tilde{S}_{\gamma, \gamma^{\prime}}(t,t^{\prime})\simeq
-i\,\int dt_{1}\int dt_{3}V_{\gamma_{1}\gamma_{2}}(t,t_{1})D_{\gamma_{2},\gamma_{3}}(t_{1},t_{3})\times \\
& \times\left( \delta_{\gamma \gamma_{1}} \delta
D_{\gamma_{3},\gamma^{\prime}}^{-1}(t_{3},t^{\prime})/\delta
U^{00}(t) +\delta
D_{\gamma_{3},\gamma^{\prime}}^{-1}(t_{3},t^{\prime})/ \delta
U^{\gamma_{1}\gamma}(t)\right)
\end{eqnarray*}

Replacing the full vertexes
\begin{equation}
\Gamma_{\xi}^{\gamma_3,\gamma^{\prime}}(t_{3},t^{\prime};t)\equiv\frac{\delta
D^{-1}_{\gamma_{3},\gamma^{\prime}}(t_{3},t^{\prime})}{\delta U^{\xi}(t)}
\label{dfVertx}%
\end{equation}
by zero ones and using Eqs.(\ref{Notat_P_Delt}),(\ref{dfD0}), we
find:
\begin{eqnarray}
&&\Gamma_{00}^{\gamma_{3},\gamma^{\prime}}(t_{3},t_{1};t) \simeq\lbrack\Gamma^{(0)}
(t_{3},t_{1};t)]_{00}^{\gamma_{3},\gamma^{\prime}}\equiv\gamma_{00}^{\gamma_{3},\gamma^{\prime}}
(t_{3},t_{1};t)\nonumber\\
&&  =\frac{\delta\lbrack D^{(0)-1}(t_{3},t_{1})]_{\gamma_{3},\gamma^{\prime}}}{\delta
U^{00}(t)}=-\delta(t_{3}-t_{1})\frac{\delta\Delta_{\gamma_{3}\gamma^{\prime}}(t_{1})}
{\delta U^{00}(t)}\nonumber\\
&&  =-\delta(t_{3}-t_{1})\frac{\delta\lbrack\Delta_{\gamma_{3}}^{(0)}
\delta_{\gamma^{\prime}\gamma_{3}}+\left[  U^{\gamma_{3}\gamma^{\prime}}(t_{1})-
\delta_{\gamma_{3}\gamma^{\prime}}U^{00}
(t_{1})\right]  ]}{\delta U^{00}(t)}\nonumber\\
&&
=\delta_{\gamma_{3}\gamma^{\prime}}\delta(t_{3}-t_{1})\delta(t-t_{1});
\label{p00}
\end{eqnarray}
\begin{eqnarray}
&&\gamma_{\gamma_{1}\gamma}^{\gamma_{3},\gamma^{\prime}}(t_{3},t_{1};t)=
-\delta(t_{3}-t_{1})\times\nonumber\\
&&\times \frac{\delta\lbrack\Delta_{\gamma_{3}}^{(0)}\delta_{\gamma^{\prime}\gamma_{3}}+
\left[U^{\gamma_{3}\gamma^{\prime}}(t_{1})-\delta_{\gamma^{\prime}\gamma_{3}}U^{00}(t_{1})\right]  ]}
{\delta U^{\gamma_{1}\gamma}(t)}\nonumber\\
&&  =-\delta_{\gamma_{3}\gamma_{1}}\delta_{\gamma \gamma^{\prime}}
\delta(t_{3}-t_{1})\delta(t-t_{1}). \label{p-pp}
\end{eqnarray}
As a result, we obtain
\begin{eqnarray}
&&\tilde{S}_{\gamma, \gamma^{\prime}}^{(1)}(t,t^{\prime})=-i\int dt_{1}V_{\gamma\gamma_{2}}
(t,t_{1})D_{\gamma_{2},\gamma^{\prime}}(t_{1},t^{+})\delta(t-t^{\prime})\nonumber\\
&&  +i\delta_{\gamma \gamma_{1}}\int
dt_{1}V_{\gamma_{1}\gamma_{2}}(t,t_{1})
D_{\gamma_{2},\gamma_{1}%
}(t_{1},t^{+})\delta(t-t^{\prime}).
\label{Sshift}
\end{eqnarray}
As seen, this contribution is indeed local in time and, therefore,
represents an effective field, which shifts the transition
energies in the quantum dot.

\section{The diagonal approximation}

The analysis of magneto-transport through a multilevel quantum
dot, even  within the MFA, is still a formidable numerical task.
Fortunately, as discussed above (see the arguments between
Eqs.(\ref{alpha}) and (\ref{diagCurr})) we can use a diagonal
approximation. This makes the problem much easier numerically and,
what is more important, it allows to get advanced quite far in
analytical treatment. The non-diagonal contributions certainly are
important for the formation of Kondo state, however, their role is
less important in the resonance-tunnelling regime.

In the diagonal approximation the  self-operator takes the form
\begin{eqnarray}
&  \tilde{S}_{\gamma,\gamma}^{(1)}(t,t^{\prime})=\delta(t-t^{\prime})
\Sigma_{0\gamma}^{shift}(t),\nonumber\\
&  \Sigma_{0\gamma}^{shift}(t)=i\int dt_{1}\sum_{\gamma_{1}}V_{\gamma_{1}\gamma_{1}}(t,t_{1})
D_{\gamma_{1},\gamma_{1}}(t_{1},t^{+})\times\nonumber\\
&\times (1-\delta_{\gamma \gamma_1})
 \label{SshiftDiag}%
\end{eqnarray}

As seen, the self-interaction is cancelled. This  nice feature of
the theory  is provided by the commutation relations. The
mechanism is as follows. We recall that in the equation of motion
for the Hubbard operator, in the first order with respect to the
tunnelling matrix elements $v$, the anticommutator
$\left\{X^{0\gamma},v^{\dagger}X^{\gamma 0}\right\}c_p  $
generates the operator
$\hat{P}^{0\gamma}=Z^{00}+Z^{\gamma\gamma}$. In order to obtain
the first order in the perturbation theory with respect to
effective interaction $V=v^{\dagger}C(t,t^{\prime})v$ we have to
calculate the second order with respect to $v$. This gives rise to
the commutator $\left[ X^{0\gamma},\hat{P}^{0\gamma}\right]
=\left[  X^{0\gamma},Z^{00}\right]  +\left[  X^{0\gamma},Z^{\gamma\gamma}\right]  =-X^{0\gamma}%
+X^{0\gamma}=0$ . Thus, indeed, the cancellation of the
self-interaction is an inner property of the theory.

In the diagonal approximation the Hubbard-I term of the  self-operator, 
Eq.(\ref{S_HIA}), is
\begin{equation}
S_{\gamma,\gamma}^{HI}(t,t_{1})=P^{\gamma\gamma}(t)\,V_{\gamma\gamma}(t,t_{1}).
\label{S_HIAdiag}
\end{equation}
Since there is only one type of coupling between transitions,
$\left[0,\gamma\right]  \Longleftrightarrow\left[\gamma,0\right]
$,  we can use the simplified notations:
\begin{eqnarray}
&S_{\gamma,\gamma}^{HI}\equiv S_{\gamma}^{HI},P^{\gamma\gamma}\equiv P_{\gamma},
V_{\gamma\gamma}\equiv V_{\gamma},\tilde{S}_{\gamma,\gamma}^{(1)}\equiv\tilde{S}_{\gamma}^{(1)}\nonumber\\
&\Delta_{\gamma\gamma}^{(0)}\equiv\Delta_{\gamma}^{(0)},\Sigma_{0\gamma}^{shift}\equiv
\Sigma_{\gamma}^{shift},D_{\gamma,\gamma}^{HI}\equiv D_{\gamma},...
\label{Notations}
\end{eqnarray}

Equation for the locator, Eq.(\ref{D_HIA}), transforms in these notations
to the form
\begin{equation}
\int dt_{1}\left\{  d_{\gamma}^{-1}(t,t_{1})-\,S_{\gamma}^{HI}(t,t_{1})\right\}
D_{\gamma}(t_{1},t^{\prime})=\delta(t-t^{\prime}), \label{LocMFA}%
\end{equation}
where
\begin{equation}
d_{\gamma}^{-1}(t,t_{1})=\left[  i\partial_{t}-\Delta_{\gamma}^{(0)}(t)-\Sigma
_{\gamma}^{shift}(t)\right]  \delta(t-t_{1}). \label{LocShift}%
\end{equation}
Equation for the GF $G_{\gamma}$ ($\equiv G_{\gamma,\gamma}$ ) can be written as follows:

\begin{eqnarray}
&  \int dt_{1}\sum_{\gamma_{1}}\Bigg{\{} \left[ i\partial_{t}
-\Delta_{\gamma}^{(0)}(t)-\Sigma_{\gamma}^{shift}(t)\right]
\delta(t-t_{1})
\nonumber\\
& - S_{\gamma}^{HI}(t,t_{1})\Bigg{\}}  G_{\gamma}(t_{1},t^{\prime}%
)=\delta(t-t^{\prime})P_{\gamma}(t^{\prime}). \label{da1}%
\end{eqnarray}
Now, when the equations for  the GFs and locators in the MFA are derived,
we can put auxiliary fields $U^{\xi}(t)=0$.  After that the
Hamiltonian does not depend on time any more and, therefore, the
locators and GFs depend only on time difference.
$P_{\gamma}(t^{\prime})$ does not depend on time as well:
$P_{\gamma}(t^{\prime})\rightarrow P_{\gamma}=N_{0}+N_{\gamma}$.
The latter simplifies the problem of derivation of the expressions
for $S_{\gamma}^{HI}(t-t_{1})$ for real times ($R$ stands for
\emph{Retarded} and $A$ for \emph{Advanced} GFs):
\begin{equation}
\left[  S_{\gamma}^{HI}(t-t_{1})\right]  ^{R,A,>,<}=P_{\gamma}\,\left[  V_{\gamma}%
(t-t_{1})\right]  ^{R,A,>,<} \label{da2}%
\end{equation}
For the transition energy shift $\Sigma_{\gamma}^{shift}$ (that
also does not depend on time at $U^{\xi}(t)=0$) the ''lesser''
value must be used. As a result, Eq.(\ref{LocMFA}), written in the
form of the integral equation on the imaginary time-axis
\begin{eqnarray}
&&D_{\gamma}(t-t^{\prime})\,=d_{\gamma}(t-t^{\prime})+\label{dac1}\\
&&\int dt_{1}\int dt_{2}d_{\gamma}(t-t_{2})
P_{\gamma}V_{\gamma}(t_{2}-t_{1})\,D_{\gamma}(t_{1}-t^{\prime}),
\nonumber
\end{eqnarray}
can be immediately continued into the real-time axis.

Correspondingly, the continued equations have the form
\begin{eqnarray}
&D_{\gamma}^{R/A}(t-t^{\prime})\,=d_{\gamma}^{R/A}(t-t^{\prime})+\label{dac2}\\
&d_{\gamma}^{R/A}(t-t_{2})P_{\gamma}V_{\gamma}^{R/A}(t_{2}-t_{1})\,D_{\gamma}^{R/A}(t_{1}-t^{\prime})
\nonumber
\end{eqnarray}
\begin{eqnarray*}
&&D_{\gamma}^{<}(t-t^{\prime})\,=d_{\gamma}^{<}(t-t^{\prime})+\nonumber\\
&&+\int_{-\infty}^{\infty}dt_{1}\int_{-\infty}^{\infty}dt_{2}
\left[  d_{\gamma}(t-t_{2})P_{\gamma}V_{\gamma}(t_{2}-t_{1})\,\right]^{<}
D_{\gamma}^{A}(t_{1}-t^{\prime})\\
&&  +\int_{-\infty}^{\infty}dt_{1}\int_{-\infty}^{\infty}dt_{2}\;d_{\gamma}%
^{R}(t-t_{2})P_{\gamma}V_{\gamma}^{R}(t_{2}-t_{1})\,D_{\gamma}^{<}(t_{1}-t^{\prime}),
\end{eqnarray*}
or, opening the square brackets,
\begin{eqnarray}
&&D_{\gamma}^{<}(t-t^{\prime})\,=d_{\gamma}^{<}(t-t^{\prime})+\label{dac3}\\
&&+\int_{-\infty}^{\infty}dt_{1}\int_{-\infty}^{\infty}
dt_{2}d_{\gamma}^{<}(t-t_{2})P_{\gamma}V_{\gamma}^{A}(t_{2}-t_{1})\,D_{\gamma}^{A}(t_{1}-t^{\prime})\nonumber\\
&&  +\int_{-\infty}^{\infty}dt_{1}\int_{-\infty}^{\infty}dt_{2}d_{\gamma}%
^{R}(t-t_{2})P_{\gamma}V_{\gamma}^{<}(t_{2}-t_{1})\,D_{\gamma}^{A}(t_{1}-t^{\prime
})\nonumber\\
&&  +\int_{-\infty}^{\infty}dt_{1}\int_{-\infty}^{\infty}dt_{2}d_{\gamma}%
^{R}(t-t_{2})P_{\gamma}V_{\gamma}^{R}(t_{2}-t_{1})\,D_{\gamma}^{<}(t_{1}-t^{\prime}).
\nonumber
\end{eqnarray}
Using the matrix form, we can simplify the latter equation
\begin{equation}
\left[  \mathbf{1}-\mathbf{d}^{R}\mathbf{PV}^{R}\,\right]  \mathbf{D}%
^{<}=\mathbf{d}^{R}\mathbf{PV}^{<}\,\mathbf{D}^{A}+\mathbf{d}^{<}%
\mathbf{PV}^{A}\,\mathbf{D}^{A}.
\label{dac4}
\end{equation}
As follows from Eq.(\ref{dac2}), the square bracket in the left-hand side of
Eq.(\ref{dac4}) is $\mathbf{d}^{R}\left[  \mathbf{D}^{R}\right]  ^{-1}$.
Multiplying Eq.(\ref{dac4}) from the left by
$\left[  \mathbf{d}^{R}\right]^{-1}$ and taking into account that
$\left[  \mathbf{d}^{R}\right]^{-1}\mathbf{d}^{R}=1$,
$\left[  \mathbf{d}^{R}\right]  ^{-1}\mathbf{d}^{<}=0$, we have
\begin{equation}
\left[  \mathbf{D}^{R}\right]  ^{-1}\mathbf{D}^{<}=\mathbf{PV}^{<}%
\,\mathbf{D}^{A} \label{dac6}%
\end{equation}
 Thus, we find the same relation which is known in the theories for fermions
 (see Ref.\onlinecite{KB}):
\begin{eqnarray}
&D_{\gamma}^{<}(t-t^{\prime})\,=\int_{-\infty}^{\infty}dt_{1}
\int_{-\infty}^{\infty}dt_{2}D_{\gamma}^{R}(t-t_{2})P_{\gamma}\times\nonumber\\
&\times V_{\gamma}^{<}(t_{2}-t_{1})\,D_{\gamma}^{A}(t_{1}-t^{\prime})
\label{dac7}
\end{eqnarray}
The advantage of this expression consists in the fact that it does
not contain bare, non-renormalized magnitudes. On the other hand,\
its validity is restricted to: i) MFA and ii) stationary states
(the step from the differential equation Eq.(\ref{dac6}) to the
integral one, Eq.(\ref{dac7}), uses the boundary condition; the
same statement is valid for Eq.(\ref{dac2})) as well. The
expression for $D_{\gamma}^{>}$, obtained in a similar fashion,
leads to Eq.(\ref{dac7}), where $V_{\gamma}^{<}$ should be
replaced by $V_{\gamma}^{>}$. Making Fourier transformation, we
find:
\begin{eqnarray}
D_{\gamma}^{R/A}(\omega)\, &=&\frac{1}{\left[  d_{\gamma}^{R/A}(\omega)\right]
^{-1}-P_{\gamma}V_{\gamma}^{R/A}(\omega)}\nonumber\\
&=& \frac{1}{\omega-\Delta_{\gamma 0}-P_{\gamma}V_{\gamma}
^{R/A}(\omega)\pm i\delta}\label{dac8}\\
D_{\gamma}^{<,>}(\omega)\,  &
=&D_{\gamma}^{R}(\omega)P_{\gamma}V_{\gamma}^{<,>}(\omega
)\,D_{\gamma}^{A}(\omega). \label{dac9}
\end{eqnarray}
Multiplying Eqs.(\ref{dac8}),(\ref{dac9}) by $P_{\gamma}$ we obtain GFs (see
Eq.(\ref{G=DP})):%
\begin{eqnarray}
G_{\gamma}^{R/A}(\omega)\,  &  =\frac{P_{\gamma}}{\omega-\Delta_{\gamma 0}-P_{\gamma}V_{\gamma}%
^{R/A}(\omega)\pm i\delta}
\label{dac8_G}\\
G_{\gamma}^{<,>}(\omega)\,  &  =G_{\gamma}^{R}(\omega)V_{\gamma}^{<,>}(\omega)\,G_{\gamma}%
^{A}(\omega).
\label{dac9_G}%
\end{eqnarray}

The effective interaction $V_{\gamma}^{>,<}$ expressed in terms of the width functions
$\Gamma_{\gamma}^{\lambda}(\omega) (\lambda=l,r$; see also Appendix \ref{app2}) is :
\begin{eqnarray}
V_{\gamma}^{>}(\omega) & = &-i\sum_{\lambda}\Gamma_{\gamma}^{\lambda}(\omega)
(1-f_{\lambda}(\omega))
\nonumber\\
V_{\gamma}^{<}(\omega)& = &i\sum_{\lambda}\Gamma_{\gamma}^{\lambda}(\omega)f_{\lambda}(\omega)
\label{da7}\\
V_{\gamma}^{R/A}(\omega)& = & \sum_{\lambda}V_{\gamma}^{\lambda,R/A}(\omega)=
\Lambda_{\gamma}(\omega)\mp i\Gamma_{\gamma}(\omega)\nonumber
\end{eqnarray}
where
\begin{eqnarray*}
\Lambda_{\gamma}(\omega)&=&\sum_{k\sigma \in \lambda,\lambda}
v_{\gamma,k\sigma}^{\lambda}\frac{\cal{P}}{\omega-\varepsilon_{\lambda
k\sigma}}v_{k\sigma,\gamma}^{\lambda},\\
\Gamma_{\gamma}(\omega) &\equiv &
\Gamma_{\gamma}^{l}(\omega)+\Gamma_{\gamma}^{r}(\omega),
\end{eqnarray*}
where $\cal{P}$ denotes the principal part of the integral. The
real part\ $\Lambda_{\gamma}(\omega)$ in the retarded and advanced
interaction $V_{\gamma}^{R/A}(\omega)$ can be neglected in a
wide-band case ($\Lambda_{\gamma}(\omega)\simeq\left[ \left\langle
\left| v_{\gamma,k\lambda\sigma}\right| ^{2}\right\rangle
/W\right] \ln\left|  \left(  W+\omega\right) /\left(
W-\omega\right) \right| $, where $W$ is of order of a half of a
bandwidth and we are interested in $\omega\ll W$ ). Then
$V_{\gamma}^{R/A}(\omega)\simeq\mp i\Gamma_{\gamma}$ and,
therefore,
\begin{equation}
G_{\gamma}^{R/A}(\omega)\,=\frac{P_{\gamma}}{\omega-\Delta_{\gamma 0}\pm iP_{\gamma}\Gamma_{\gamma}}.
\label{G_R/A}%
\end{equation}
Using this equation, we obtain a few useful relations that will be used below.
In particular, the difference of GFs is
\begin{equation}
G_{\gamma}^{R}(\omega)-G_{\gamma}^{A}(\omega)=-2\pi iP_{\gamma}L_{\gamma}(\omega),
\label{diffGRGA}
\end{equation}
where
\begin{equation}
L_{\gamma}(\omega)\equiv\frac{P_{\gamma}\Gamma_{\gamma}/\pi}{\left(  \omega-\Delta_{\gamma}\right)^{2}
+\left(  P_{\gamma}\Gamma_{\gamma}\right)^{2}}
\label{Lorentz}
\end{equation}
is the Lorentz distribution.  Note that, contrary to the WCI case, in
the SCI case the width of the distribution  depends on the
non-equilibrium population numbers via $P_{\gamma }=N_{0
}+N_{\gamma }$. With the aid of Eqs.(\ref{da7}),(\ref{G_R/A}), we
obtain for the GF $G^{<}$:
\begin{equation}
G_{\gamma}^{<}(\omega)=G_{\gamma}^{R}(\omega)V_{\gamma}^{<}(\omega)\,G_{\gamma}^{A}(\omega)
=2\pi iP_{\gamma}L_{\gamma}(\omega)\bar{f}(\omega), \label{Glesser}%
\end{equation}
where $\alpha_{\gamma}^{l/r}\equiv\Gamma_{\gamma}^{l/r}/\Gamma_{\gamma}$ and $\bar{f}%
(\omega)$ is the weighted Fermi-function:
\begin{equation}
\bar{f}_{\gamma}(\omega)\equiv\left[  \alpha_{\gamma}^{l}f_{l}(\omega)+\alpha_{\gamma}%
^{r}f_{r}(\omega)\right]  . \label{weightedFermi}%
\end{equation}

\subsection{Equations for population numbers.}

All  the locators and GFs depend on the single-time correlators $P_{\gamma}%
\equiv\left\langle X^{0\gamma}X^{\gamma 0}+X^{\gamma 0}X^{0\gamma}\right\rangle =N_{0}+N_{\gamma}$. The
expectation values $\left\langle X^{\gamma 0}X^{0\gamma}\right\rangle $ and $\left\langle
X^{0\gamma}X^{\gamma 0}\right\rangle $ can be found by means of GFs $G_{\gamma}^{<,>}(t,t)$.
We recall that our aproach is developed for the stationary case only. Thus, we have
\begin{eqnarray}
&&\left\langle X^{\gamma 0}X^{0\gamma}\right\rangle =-i(i\left\langle X^{\gamma 0}(t)X^{0\gamma}%
(t)\right\rangle )=\nonumber\\
&&=-iG_{\gamma}^{<}(t,t^{+})= -i\int d\omega G_{\gamma}^{<}(\omega)
\label{PopNumb1}
\end{eqnarray}
Since the GFs depend on $P_{\gamma}$ themselves, Eqs.(\ref{PopNumb1}) for the
population numbers are, in fact, the equations of self-consistency:
\begin{eqnarray}
N_{\gamma}  &  =-i\int d\omega G_{\gamma}^{<}(\omega)\nonumber\\
N_{0}  &  =i\int d\omega G_{\gamma}^{>}(\omega)
\label{EqForPopN}
\end{eqnarray}
The normalization condition, which follows from the averaging
with respect to the state/ensemble of interest of Eq.(\ref{UnityExpansion}),
provides the additional equation
\begin{equation}
1=N_{0}+\sum_{\gamma}N_{\gamma}
\label{norm}%
\end{equation}

Substituting Eq.(\ref{Glesser}) into Eq.(\ref{EqForPopN}) we obtain
\begin{equation}
N_{\gamma}=P_{\gamma}s_{\gamma},\;
s_{\gamma}=\int d\omega L_{\gamma}(\omega)\bar{f}_{\gamma}(\omega).
\label{N_p}
\end{equation}
From Eqs.(\ref{dac9}),(\ref{Glesser}),(\ref{N_p}) one may conclude that the magnitude
$s_{\gamma}$ is the integrated ''lesser'' part of the locator:%
\begin{equation}
D_{\gamma}^{<}(\omega)=2\pi iL_{\gamma}(\omega)\bar{f}(\omega). \label{Dless}%
\end{equation}
For $\Gamma\beta \ll1$, the Lorentzian can be approximately
treated as a delta-function. Therefore, the integral in Eq.(\ref{EqForPopN}) can be
approximated by values of the weighted Fermi function, Eq.(\ref{weightedFermi}),
at the transition energy. It results in a simple expression:
\begin{equation}
N_{\gamma}=P_{\gamma}\tilde{s}_{\gamma},\;\tilde{s}_{\gamma}
\simeq\left[f_{l}(\Delta_{\gamma})\alpha_{\gamma}^{l}+
f_{r}(\Delta_{\gamma})\alpha_{\gamma}^{r}\right].
\label{N_p_lowGamma}
\end{equation}
In numerous experiments \cite{kou,tr,ihn,ota} with  quantum dots
the temperature is usually lower than the level spacing. At this condition,
if the transition energy $\Delta_{\gamma}$ is above both, left and  right
electrochemical potentials, $s_{\gamma}$ is exponentially small in
Eq.(\ref{N_p_lowGamma}). This makes $N_{\gamma}$ exponentially small as well. If,
\emph{vice versa}, $\Delta_{\gamma}$ is below both $\mu_{l,r}$, $s_{\gamma}\simeq1$,
and we obtain $N_{0}\ll1$. \ Thus, the only case, when $N_0$ and $N_p$ are finite, is
when $\Delta_{\gamma}\ $\ is located within the ''conducting window'' (CW), i.e.,
in the interval of energies between $\mu_{r}$ and $\mu_{l}$.
\ This conclusion  is valid, of course, for a finite width $\Gamma_{\gamma}$
of transitions $\Delta_{\gamma}$, when $s_{\gamma}$ is defined by Eq.(\ref{N_p}).

Let us introduce the function
\begin{equation}
\Phi_{\gamma}\equiv(s_{\gamma}^{-1}-1)^{-1}
\label{Phi}
\end{equation}
that couples the population numbers of \ empty and single-electron
states:
\begin{equation}
N_{\gamma}=N_{0}\Phi_{\gamma}. \label{Np=N0_Phi}%
\end{equation}
The normalization condition yields
\begin{equation}
N_{0}=\frac{1}{1+\sum_{\gamma}\Phi_{\gamma}},\;
N_{\gamma}=\frac{\Phi_{\gamma}}{1+\sum_{\gamma}\Phi_{\gamma}}.
\label{Np_and_N0}%
\end{equation}

In order to display highly non-linear
dependence of the population numbers on the applied voltage,
let us consider the limit $s_{\gamma}\rightarrow\tilde
{s}_{\gamma}$.
As a result, we obtain
\begin{eqnarray}
&&\Phi_{\gamma}\rightarrow\tilde{\Phi}_{\gamma}=
e^{-\beta\left(  \Delta_{\gamma}-\mu\right)}\times\label{PhiLimit}\\
&&\times\frac{1+e^{\beta\left(  \Delta_{\gamma}-\mu\right)  }\left[  \cosh\left(  \beta
eV/2\right)  -\delta\alpha\sinh\left(  \beta eV/2\right)  \right]  }
{e^{\beta\left(  \Delta_{\gamma}-\mu\right)  }+\cosh\left(  \beta eV/2\right)
+\delta\alpha\sinh\left(  \beta eV/2\right)},\nonumber
\end{eqnarray}
where we have introduced the following notations:
$\mu=(\mu_{l}+$ $\mu_{r})/2$ , \ $eV=\mu_{l}-$ $\mu_{r}$  and
the degree of asymmetry of the contacts $\delta\alpha=\alpha_{\gamma}^{l}-\alpha_{\gamma}^{r}$.

At $eV=0$ we have $\tilde{\Phi}_{\gamma}=\exp\left\{  -\beta\left(
\Delta_{\gamma}-\mu\right)  \right\}  $ and, therefore, the Gibbs-ensemble limit holds:
\begin{equation}
N_{0}=\frac{1}{1+\sum_{\gamma}e^{-\beta\left(  \Delta_{\gamma}-\mu\right)  }},
\;N_{\gamma}=\frac{e^{-\beta\left(  \Delta_{\gamma}-\mu\right)  }}{1+\sum_{\gamma}%
e^{-\beta\left(  \Delta_{\gamma}-\mu\right)  }}.
\label{GibbsLim}
\end{equation}
In the limit of a small bias voltage, $\beta eV/2\ll1$, we obtain
\begin{equation}
\tilde{\Phi}_{\gamma}\approx e^{-\beta\left(  \Delta_{\gamma}-\mu\right)  }\left[
1+\delta\alpha\frac{\beta eV}{2}\right].
\label{small_eV}
\end{equation}
The most prominent phenomenon occurs for a symmetric coupling to the contacts,
$\delta\alpha=0$. Namely, in this case we have
\begin{equation}
\tilde{\Phi}_{\gamma}=e^{-\beta\left(  \Delta_{\gamma}-\mu\right)
}\frac{1+e^{\beta\left(  \Delta_{\gamma}-\mu\right)  }\cosh\left(  \beta
eV/2\right)  }{e^{\beta\left(  \Delta_{\gamma}-\mu\right)  }+\cosh\left(  \beta
eV/2\right)  }. \label{symmPhi}%
\end{equation}
When a ''level'' $\Delta_{\gamma}$ gets into the CW, $\delta eV=\left\{  \mu
_{r}<\Delta_{\gamma}-\mu<\mu_{l}\right\}  $, $\exp\left\{  \beta\left(
\Delta_{\gamma}-\mu+eV/2\right)  \right\}  \gg1$. At large bias voltage
$\tilde{\Phi}_{\gamma}\rightarrow1$  and all population numbers become equal to
each other
\begin{equation}
N_{0}=N_\gamma=\frac{1}{1+\sum_{\gamma\in\delta eV}1}
\label{EqualPopN}%
\end{equation}
Population numbers of the states with the energies outside
of the interval $\delta eV$ are equal to zero.  Thus, we find a remarkable
feature of the correlated transport for a symmetric-coupling design of the device:
\emph{large enough bias voltage equalizes the population of conducting
states.} We will return to this feature later.

\subsection{The real-time equation for energy shifts.}

We have to define the renormalized transition energy $\Delta_{\gamma}$ which is
still unknown.
The transiton-energy shift
(\ref{SshiftDiag}) contains the integral
\begin{equation}
\tilde{l}_{\gamma}=i\int_{t_{0}}^{t_{0}-i\beta}dt_{1}V_{\gamma}(t,t_{1})
D_{\gamma}(t_{1},t^{+}),
\label{da3}
\end{equation}
The continued to real time expression, corresponding to Eq. (\ref{da3}), will be
denoted as $l_{\gamma}$. We have
\begin{equation}
\Sigma_{\gamma}^{shift}=\sum_{\gamma_{1}}l_{\gamma_{1}}-l_{\gamma},
\label{da4}
\end{equation}
where (see Ref.\onlinecite{KB})
\begin{eqnarray}
&&l_{\gamma}= \int_{-\infty}^{t}d\tau\left\{  V_{\gamma}^{>}(t,\tau)D_{\gamma}^{<}%
(\tau,t)-V_{\gamma}^{<}(t,\tau)D_{\gamma}^{>}(\tau,t)\right\} \nonumber\\
&&=\int_{-\infty}^{\infty}\frac{dE}{2\pi i}\frac{\Pi_{\gamma}(E)}{E+i\delta},
\label{da5}%
\end{eqnarray}
and
\begin{equation}
\Pi_{\gamma}(E)\equiv\int_{-\infty}^{\infty}\frac{d\omega}{2\pi}
\left\{  V_{\gamma}^{>}(\omega)D_{\gamma}^{<}(\omega+E)-
V_{\gamma}^{<}(\omega)D_{\gamma}^{>}(\omega+E)\right\}
\label{da6}%
\end{equation}
Since we assume the adiabatic switching of the interaction and,
therefore, ignore the contribution from the part of the contour $\left(
t_{0},t_{0}-i\beta\right)  $ at $t_{0}\rightarrow-\infty$, the equivalent
forms of this expression can be obtained by either consideration of the
integral (\ref{da3}) on the Keldysh contour, or by application of the
Langreth' rules (in this case $\lim_{t^{\prime}\rightarrow t+0}\left[
V_{\gamma}(t,t_{1})D_{\gamma}(t_{1},t^{\prime})\right]  ^{<}=\lim_{t^{\prime}\rightarrow
t+0}\left[  V_{\gamma}^{<}(t,t_{1})D_{\gamma}^{A}(t_{1},t^{\prime})+V_{\gamma}^{R}%
(t,t_{1})D_{\gamma}^{<}(t_{1},t^{\prime})\right]  $ ).

Thus,  substituting Eqs.(\ref{da7}) into Eq.(\ref{da6}), we obtain
\begin{eqnarray}
& & l_{\gamma}=\int_{-\infty}^{\infty}
\frac{dE}{2\pi i}\frac{\Pi_{\gamma}(E)}{E+i\delta}\nonumber\\
&&=-i\sum_{\lambda}\int_{-\infty}^{\infty}
\frac{d\omega}{2\pi}\Gamma_{\gamma}^{\lambda}(\omega)
\Bigg{\{}  (  1-f_{\lambda}(\omega))  \int_{-\infty}^{\infty}\frac{dE}{2\pi i}
\frac{D_{\gamma}^{<}(E)}{E-\omega+i\delta}\nonumber\\
&&+f_{\lambda}(\omega)\int_{-\infty}^{\infty}\frac{dE}{2\pi i}
\frac{D_{\gamma}^{>}(E)}{E-\omega+i\delta}\Bigg{\}}
\label{da9}
\end{eqnarray}
The GFs $D_{\gamma}^{<}$ have a physical meaning of distribution
functions. Since we consider wide conduction bands in contacts,
the width functions $\Gamma$ are replaced by a constant  and
extracted from the integral in Eq.(\ref{da9}). In fact, the
 limits of the integration are determined by the bottom and the top of
conduction bands (''left'' and ''right''). It follows from the
definition of the width, Eq.(\ref{width}):
$\Gamma_{\gamma}^{\lambda}(\omega)\neq0$ only within the interval
of $\omega$, where the density of states of the corresponding
conduction band is nonzero.  The remaining integral describes the
distribution of the density of the mixed charge ($\left\langle
c^{\dagger}X\right\rangle $) over the energy, which is responsible
for the renormalization of the transition energy
$\Delta_{\gamma}$.

Inserting into Eq.(\ref{da9}) the expressions for "lesser" and
"greater" locator, which within our approximation are $D^{<}=2\pi
iL_{\gamma}\bar{f}$ and $D^{>}=-2\pi iL_{\gamma}(1-\bar{f})$, we
obtain  a simple formula for the partial shift of the transition
energy
\begin{equation}
l_{\gamma}=\frac{1}{\pi}\int_{-W}^{W}d\omega
\int_{-W}^{W}dE\;L_{\gamma}(E)
\frac{\Gamma_{\gamma}(\omega)[\bar{f}(E)-\bar{f}(\omega)]}{E-\omega+i\delta}
\label{da12}%
\end{equation}
Here, $W$ is a half of bandwidth, which is the same in the both
contacts. In fact, the integral limits can be replaced by
infinities, since the bandwidth is much bigger than any other
parameter in the theory.

The right-hand side of Eq.(\ref{da9}) contains dressed locators $D_{\gamma}^{<,>}$.
Therefore, the system of equations
\begin{eqnarray}
\Delta_{\gamma}  & = &\Delta_{\gamma}^{(0)}+\Sigma_{\gamma}^{\it shift},\nonumber\\
\Sigma_{\gamma}^{\it shift}  & = &\sum_{\gamma_{1}}l_{\gamma_{1}}-l_{\gamma},
\label{SystForDelt}%
\end{eqnarray}
and Eq.(\ref{da9}) have to be solved self-consistently.

There are a few remarks in order. First, the Lorentzian width, Eq.(\ref{Lorentz}),
is determined by the product $P_{\gamma}\Gamma_{\gamma}=\left(N_{0}+N_{\gamma}\right)\Gamma_{\gamma}$.
Therefore, if the transition energy $\Delta_{\gamma}$ is outside
of the resonant CW, the population numbers are almost equal to zero.
In this case, the coupling between the QD and contacts does not affect the discrete
transitions (their widths are zeros).  Second, the imaginary part of Eq.(\ref{da12})
is exactly equal to zero. Thus, this expression provides a simple shift
of the transition energy (as it should be). Third, using the definition  of the
width $\Gamma_p$, Eq.(\ref{width}), one can write Eq.(\ref{da12}) in the form

\begin{eqnarray}
l_{\gamma} &=&\frac{1}{2}\sum_{k\sigma \in \lambda,\lambda}
\left|  v_{k\sigma,\gamma}^{\lambda}\right| ^{2}\int_{-\infty}^{\infty}dE\;
\frac{f_{\lambda}(E)-f(\varepsilon_{\lambda k\sigma})}
{E-\varepsilon_{\lambda k\sigma}}\times\nonumber\\
&&\times\left(  -\frac{1}{\pi}Im D_{\gamma}^{R}(E)\right),
\label{K2}%
\end{eqnarray}
which reveals the other aspect of the shift.
Namely, estimating the integral at low temperatures, we obtain
\begin{eqnarray}
&&l_{\gamma}\simeq-\frac{ln R}{4W}(| v_{0,\gamma}^{l}|^2+
|v_{0,\gamma}^{r}|^2)\label{log}\\
&&R=\frac{W^{2}}{\sqrt{\left[\left(\Delta_{\gamma}-\mu_{l}\right)^{2}+\left(
\Gamma_{\gamma}P_{\gamma}\right)  ^{2}\right]  \left[  \left(  \Delta_{\gamma}-
\mu_{r}\right)^{2}+\left(\Gamma_{\gamma}P_{\gamma}\right)  ^{2}\right]}}\nonumber
\end{eqnarray}
Here, we introduced the mixing
$\left|v_{0,\gamma}^{\lambda}\right|  ^{2}\equiv\sum_{k\sigma \in
\lambda} \left|  v_{k\sigma,\gamma}^{\lambda}\right|  ^{2}$  and a
weighted density of states
$g_{\lambda}^{0}(\omega)\equiv\sum_{k\sigma \in \lambda, \lambda}
\left( \left|v_{k\sigma,\gamma}^{\lambda}\right| ^{2}/\left|
v_{0,\gamma}^{i}\right|
^{2}\right)\delta(\omega-\varepsilon_{\lambda k\sigma})$.
Evidently, that the smallness of the coupling constant $\left|
v_{0,\gamma}^{\lambda}\right|  /2W$ may be compensated by a large
logarithm, when the transition energy is in the proximity of one
of the electrochemical potentials. On the one hand, it may be
strong and, moreover, it is sensitive to the bias voltage, since
$\mu_{l/r}=\mu \pm eV/2$. On the other hand, the effect is
logarithmically weak and, besides, the infra-red cut-off in the
integral is $max\{ \Gamma P,T\} $, i.e., with  an increase of
the temperature  the renormalization becomes less effective. However,
 the effect is quite appreciable numerically (see below). Thus, we
conclude that the attachment of contacts to QD shifts the
transition energies logarithmically via formation of mixed charge
$\left\langle c^{\dagger}X\right\rangle $ in each available
channel $\gamma $.

\subsection{ Current in the diagonal approximation}

Now we have all ingredients for calculation of the current. The
final expression for the current is obtained by the substitution
Eqs.(\ref{diffGRGA}),(\ref{Glesser}), (\ref{weightedFermi}) into
Eq.(\ref{diagCurr}):
\begin{equation}
J_{l}=\frac{4\pi e}{\hbar} \sum_{\gamma}P_{\gamma}\int d\omega
\Gamma_{\gamma}\alpha_{\gamma}^{l}\alpha_{\gamma}^{r} \lbrack
f_{l}(\omega)-f_{r}(\omega)]L_{\gamma}(\omega). \label{curr1}
\end{equation}
This expression is almost identical to the one of the
non-interacting electron problem. The only difference is that the
expression for  the SCI regime contains the effective width $\Gamma P
$, where $P_{\gamma}=N_{0}+N_{\gamma}$ is a combination of
population numbers. All $P_{\gamma}<1$ due to the normalization
condition, Eq. (\ref{norm}). At $\Gamma\beta\ll1$ the integral can
be estimated by replacing the Lorentzian with the delta-function.
As a result, we obtain an analogue of the Buttiker's formula for
the strong Coulomb interaction regime
\begin{equation}
J=\frac{4\pi e}{\hbar}\sum_{\gamma}P_{\gamma}\Gamma_{\gamma}\alpha_{\gamma}^{l}\alpha_{\gamma}^{r}%
\frac{\sinh\left[  eV/2T\right]  }{\cosh\left[  \left(
\Delta_{\gamma}-\mu\right)
/T\right]  +\cosh\left[  eV/2T\right]  }.\label{Buttiker}%
\end{equation}
Here, the coefficients $\alpha_{\gamma}^{l/r}$ determine a
relative transparency of the ''left''/''right'' junction. The
degree of the channel opening is defined by the product
$\Gamma_{\gamma}\alpha_{\gamma}^{l}\alpha_{\gamma}^{r}P_{\gamma}$.
The population of each ''level'' $\Delta_{\gamma}$ depends on the
populations of other levels due to the sum rule, Eq.(\ref{norm}).
Below we apply our theory to describe the magneto-transport
through a small vertical quantum dot under a perpendicular
magnetic field.

\section{Parabolic quantum dot in magnetic field}

Shell effects are among the most remarkable phenomena observed in
vertical quantum dots \cite{tar}. In virtue of the potential
symmetries, the orbital motion of electrons could lead to the
degeneracies even at strong intra-dot Coulomb interaction (see,
for example, a discussion about hidden symmetries for a parabolic
potential in Ref.\onlinecite{sim}). Numerous papers devoted to the
analysis of nonlinear transport through QDs in magnetic field  are
focused, however, on the effect of spin splitting of levels in
 the dot. Simple estimations evidently indicate that the orbital effects
are much stronger than spin effects and as yet are not discussed
in literature related to quantum transport. For example, in a
two-dimensional harmonic oscillator model for QDs upon the
perpendicular magnetic field (cf
Ref.\onlinecite{HN}) the effective spin magnetic moment is $\mu^{\ast}%
=g_{L}\mu_{B}$ with $\mu_{B}=|e|\hbar/2m_{e}c$ and the effective Lande factor
$g_{L}=0.44$. For GaAs the effective mass $m^{\ast}=0.067m_{e}$ determines the
orbital magnetic moment for electrons and gives $\mu_{B}^{\mathrm{eff}}%
\approx15\mu_{B}$,  which is 30 times stronger than the spin one. 
In this section we discuss the effects of the
orbital motion on the nonlinear (with respect to the bias voltage)
transport in both  the WCI and SCI regimes.

 We consider a dot with a circular
shape, $\omega _{x}=\omega _{y}=\omega _{0}$ in a perpendicular
magnetic field $B$ (Ref.\onlinecite{HN}). The dot eigenmodes are
$\Omega _{\pm }=(\Omega \pm\omega _{c}/2)$ with $\Omega
=\sqrt{\omega _{0}^{2}+\omega _{c}^{2}/4}$. Here $\omega
_{c}=\frac{|e|}{m^{\ast }c}B$. We choose the position of the QD
potential well so that the first level in  the QD is above $\mu$. In
this case at zero bias voltage the QD is empty.

\begin{figure}[ht]
\centerline{\psfig{figure=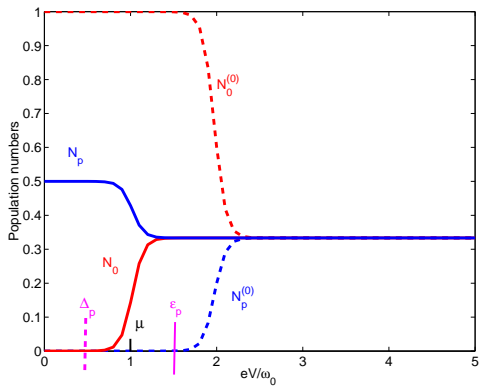,width=2.5in,clip=}}
\caption{(Color online) Population numbers $N_0, N_p$ for three
states: empty state $|0)$ and two degenerate ones $|p_1),|p_2)$, -
as a function of the bias voltage. The energies are given in units
of confinement energy $\omega_0$. The electrochemical
potential is chosen at $\mu/\omega_0=1.0$ and
$T/\omega_0=0.033$. The solid (dashed) line displays the
behavior of the population number with the renormalized (bare)
energy.} \label{1stFig}
\end{figure}

As was shown above, the coupling to contacts
leads to the shift of the transition energies and non-Gibbs behavior
of the population numbers of QD states. In addition, the
spectrum of QD in a magnetic field displays degeneracies.  Let us
illustrate these features in a simple example of two degenerate
transitions $|0) \rightarrow |p_1),|p_2)$. The solution of
self-consistent equations for this case is displayed on
Fig.\ref{1stFig}. The QD with bare energies is empty
($N_0=1$)until $eV$ riches the bare level $\epsilon_p/\omega_0
=1.5$ that is above the Fermi energy without the shift. However,
the renormalized single-electron states  are filled at much
smaller values of the bias voltage (see solid lines), since the
corresponding transition energies ($\Delta_{\gamma 0}/\omega_0
=0.5$) are below the chemical potential. One observes also that
the population numbers are equalized (=1/3) after the voltage
riches the corresponding transition energy, in the both cases.
Below we will demonstrate how this mechanism shows up in the
transport properties.

\subsection{Shell effects }

The response of the QD to applied bias voltage is determined,
evidently, by a electronic levels within the conducting window
$(\mu_l,\mu_r )$. The number of levels in the window depends : i)
on the relative position of the bottom of the confinement
potential in QD and the equilibrium position of the
electrochemical potential ($eV=0$); ii) on the strength of
magnetic field. Although the transport properties in the WCI
regime are known, we shortly summarize the corresponding effects
in order to compare it with the ones in the SCI regime.

The formula for the current for the WCI regime has the same form
as Eq.(\ref{curr1}), but contains normal width $\Gamma $ instead
of effective one $\Gamma P$. Thus, putting formally $P=1$ in
Eq.(\ref{curr1}), we obtain:

\begin{equation}
J^{WCI}=\frac{4e}{\hbar} \sum_{\gamma}\int d\omega [f_{l}(\omega
)-f_{r}(\omega)]\frac{\Gamma_{\gamma}^{l}(\omega)\Gamma_{\gamma}^{r}(\omega
)}{[\omega-\varepsilon_{\gamma}]^{2}+[\Gamma_{\gamma}(\omega)]^{2}}.
\label{WCIcurr}%
\end{equation}
In the wide-band case, as mentioned above, the width does not
depend on energy,
$\Gamma_{\gamma}^{l/r}(\omega)\rightarrow\Gamma_{\gamma}^{l/r}$.
Then at small dot-lead coupling we find:
\begin{equation}
J_{l}\simeq\frac{4\pi e}{\hbar}\sum_{\gamma}\frac{\Gamma_{\gamma}^{l}%
\Gamma_{\gamma}^{r}}{\Gamma_{\gamma}}[f_{l}(\varepsilon_{\gamma}%
)-f_{r}(\varepsilon_{\gamma})]. \label{WCIcurrSimple}%
\end{equation}

\begin{figure}[ht]
\centerline{\psfig{figure=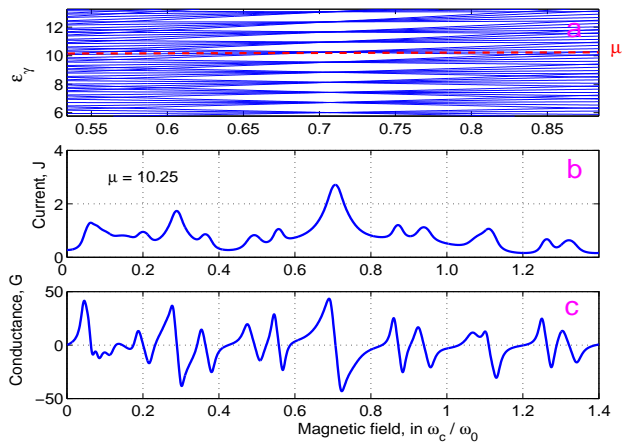,width=3.3in,height=2.3in,clip=}}
\caption{(Color online) Magnetic field dependence
($t=\omega_{c}/\omega_{0}$) of: a) a Fock-Darwin spectrum; the
position of the chemical potential $\mu$ is displayed by a dashed
line; b)the tunnelling current $J$ through the quantum dot; c)the
conductance G. See text for the definition of $J$ and $G$. All
energies are given in units of the confinement energy
$\omega_0$} \label{2dFig}
\end{figure}

The shell structure in the spectrum of a QD is evidently observed in the
WCI regime (see Fig.\ref{2dFig}a). We use a fixed value for the
 confinement energy $\omega_{0}$. The validity of
the assumption about the independence of this energy on the number
of electrons in the dot, actually, is not obvious at all. We
believe, however, that  strong enough external confining potential
and a small number of electrons (i.e., $\leq20$) all results would
persist, since the magnetic field contributes additionally to the
external confining potential.

\begin{figure}[ht]
\centerline{\psfig{figure=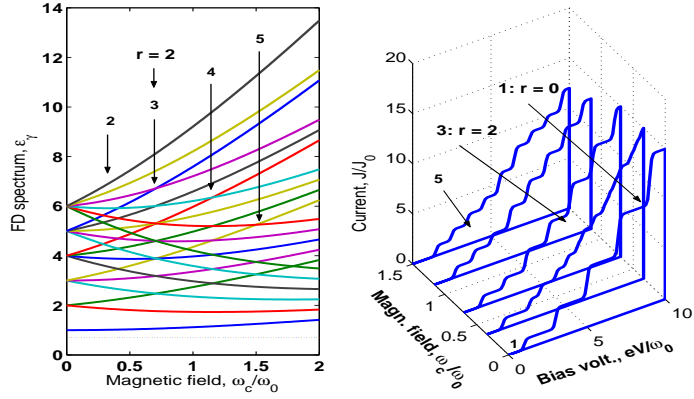,width=3.7in,height=2.2in,clip=}}
\caption{(Color online) Magnetic field
($\protect\omega_{c}/\protect\omega_{0}$) dependence of: the
Fock-Darwin spectrum (in units $\protect\omega_{0}$) (left); the
tunnelling current $J_{WCI}$ (in units $J_{0}=\bar{\Gamma}/4h$)
through the quantum dot (right). Arrows indicate the cuts in the
planes that correspond to the fields
$\protect\omega_{c}/\protect\omega_{0}=0.0, 0.34,0.7 $,
respectively. The degeneracy at
$\protect\omega_{c}/\protect\omega_{0}=0.0,0.7$ is clearly seen in
jumps of the current at $T=0$.} \label{3dFig}
\end{figure}

At given $\mu=(\mu_{L}+\mu_{R})/2$ (that simulates in our case the
choice of the gate voltage $V_{gate}$) the number of conducting
channels is determined by the number of the Fock-Darwin levels in
the conducting window $\mu_{R}<\Delta E<\mu_{L}$ (CW). This number
is changing when the magnetic field grows: the levels with higher
values of orbital (and spin) momenta $m$ move down faster than
those with lower momenta and, therefore, the high-lying levels
with large values of $L_{z}=m$ should unavoidably show up in the
CW at large enough magnetic field. The levels with the negative
orbital momenta $L_{z}=-m^{\prime}$ move up and leave the CW,
decreasing, thus, the number of conducting channels. These two
processes result in oscillations in the current (see
Fig.\ref{2dFig}b). Thus, each new level entering the CW determined
by $\mu-eV/2<\varepsilon_{\gamma}<\mu+eV/2$,
produces step in the current with the height $J_{0,\gamma}=4\pi e\Gamma_{\gamma}^{l}%
\Gamma_{\gamma}^{r}/\left( \hbar\Gamma_{\gamma}\right)  $. In a
narrow energy interval the widths $\Gamma_{\gamma}$ differ from
each other only slightly, $J_{0,\gamma} \approx J_{0}=\pi
e\Gamma/\hbar $, then the ''reduced'' current $J/J_{0}$ should
display integer steps. If some level $\varepsilon_{\gamma}$ is
$n_{\gamma}$ times degenerate, the step increases $n_{\gamma}$
times. The effect of degeneracy of the spectrum becomes
transparent at the special values of the magnetic field
\begin{equation}
t_{0}\equiv\omega_{c}/\omega_{0}=(k-1)/\sqrt{k},
\label{to}
\end{equation}
which are determined by the ratio
$\Omega_{+}/\Omega_{-}=k=1,2,3,\ldots$ of the eigen modes of the
two-dimensional quantum dot \cite{HN} (see Fig.\ref{3dFig}).

At these values the current drastically increases, whereas for the
values which are slightly larger than $t_{0}$, the negative
differential conductance, $G=dJ/dV$, arises. This is illustrated
in fig.\ref{2dFig}c.

\begin{figure}[ht]
\centerline{\psfig{figure=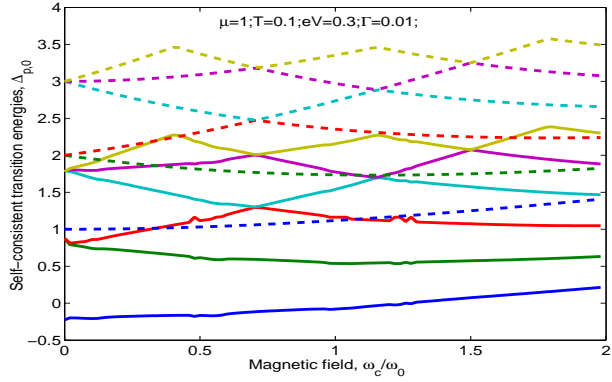,width=3.2in,height=2.0in,clip=}}
\caption{(Color online) The bare (dashed line) and renormalized
(solid line) energies as a function of the magnetic field. The
parameters of calculations are the same for the WCI and SCI
regimes. Five lowest level of the Fock-Darwin spectrum were taken
as the bare energies.} \label{4thFig}
\end{figure}

\begin{figure}[ht]
\centerline{\psfig{figure=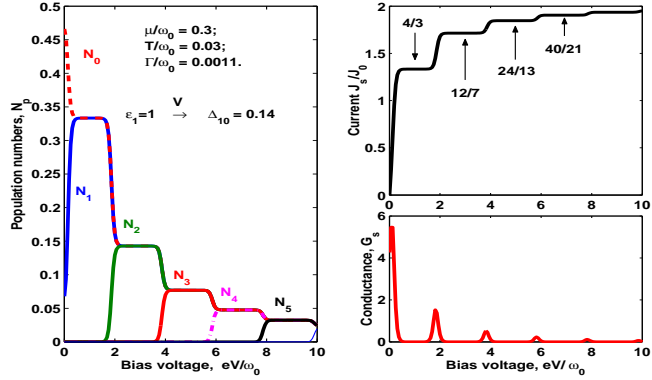,width=3.4in,height=2.0in,clip=}}
\caption{(Color online) Population numbers $N_p$ (left panel),
the current $J_S/J_0$ and the conductance $G_S=dJ_S /d(eV)$ as a function of
the bias voltage (right panel) in the SCI regime at zero magnetic field.
In each shell $k$ we have $2(k+1)$ degenerate orbitals characterized by the same
$N_p$ ($p=k+1$). In particular,  $N_0=<X^{00}>$,
$N_1=N_{k=0,\gamma _0=1}=N_{k=0,\gamma _0=2}$ {\it etc},
where $ \gamma _k$ is the orbital index in the shell $k$.
Arrows at the bias voltage axis indicate the position of the bare $\varepsilon_1$
and the renormalized $\Delta_{10}$ energies from
the shell $k=0$. Their exact values and
the parameters of the calculations are displayed on the left panel.
On the right panel, the rational numbers characterize the height of
the $n$-th step in the SCI current,
$J_{SCI}/J_{0}=2(n+1)(n+2)/(n^{2}+3n+3)$, for the last filled shell $n$.}
\label{5thFig}
\end{figure}

Fig.\ref{4thFig} displays the renormalized spectrum, obtained from
the self-consistent solution of the system of
Eqs.(\ref{SystForDelt}),(\ref{K2}). As expected from the
estimation (\ref{log}), the renormalization is quite large.  Due
to a large number of the transitions, contributing into the energy
shift (see Eq.(\ref{K2})), all transition energies are shifted more
or less homogeneously. Thus,  in spirit, this is, indeed, the MFA: the
more transitions contribute into renormalization  the closer is 
the shift of each level to an average value (mean field) (see
Fig.\ref{4thFig}).  The most prominent feature caused by these shifts
of the transition energies is a decrease of the threshold of the
bias voltage to observe the non-zero current. At small $eV$ and
chosen fixed $\mu$ (or the gate voltage) the first transition
$\Delta_{10}<\mu_{r}$, and the current, Eq.(\ref{curr1}), is zero,
in spite of $P_{0\gamma}\neq0$ ($N_{0}=1$) (see
Fig.\ref{5thFig}). At higher voltages the CW contains $n_{W}$
electron states and, according to Eq.(\ref{Np_and_N0}),
$P_{0\gamma}=2/(n_{W}+1)$. As a result, the SCI current is
$J_{SCI}=2J_{0}n_{W}/(n_{W}+1)$, where $J_{0}=\pi
e\sum_{\gamma}\Gamma _{\gamma}/\left[\hbar n_{W}\right] \equiv \pi
e\tilde{\Gamma}/\left( \hbar\right)  $. The WCI current, however,
is $J_{WCI}=J_{0}n_{W}$. Thus, even for a large bias voltage $eV$,
the SCI current is weaker than the WCI one, by factor of
$\eta=J_{SCI}/J_{WCI}=2/(n_{W}+1)$ (until the other charge sector
is not switched on at $eV\sim U$).

Let us compare the manifestation of shell degeneracies defined by the condition,
Eq.(\ref{to}), in nonlinear transport at the WCI and SCI regimes.
We consider first $r=1$, i.e., a zero magnetic field (see Fig.\ref{3dFig}).
In this case, each shell $k$ has the degeneracy $g_{k}=k+1$. If in the transport
window the last shell $n$ is filled, the total number of states involved into the
transport is $n_{W}+1=2\sum_{k=0}^{n}(k+1)+1=n^{2}+3n+3$ (2 is due to the
spin degeneracy). Consequently, the height of the $n$-th step in the SCI current is
$J_{SCI}/J_{0}=2(n+1)(n+2)/(n^{2}+3n+3)$, which is smaller than the WCI
current by factor $\eta=2/(n^{2}+3n+3)$. Since $\Phi_{\gamma} \simeq e^{-\beta
\left(  \Delta_{\gamma 0}-\mu\right)}\left[  1+\alpha\left(  eV\right)
^{2}\right]  $ at small $eV$ (see Eqs.(\ref{PhiLimit})), these effects can not be
seen in the linear conductance. Another effect (which is not seen in the
master-equation approach) follows from Eq.(\ref{SystForDelt})(see also
Fig.\ref{1stFig}): the coupling pushes
the transition energies $\Delta_{\gamma 0}$ down compared to the bare energies
$\varepsilon_{\gamma}$, which decreases the bias voltage threshold, as was mentioned
above, for the current to be observed. At $r=2$ ($\omega_{c}/\omega_{0}\simeq0.7$),
a new shell structure (see Fig.\ref{3dFig})
arises \emph{as if} the confining potential would be a deformed harmonic oscillator
without magnetic field. The number of levels are just the number of levels
obtained from the 2D oscillator with $\omega_{>}=2\omega_{<}$ ($\omega_{>}$
and $\omega_{<}$ denote the larger and smaller value of the two frequencies).
In this case $n_{W}=(n+2)^{2}/2$ if the last filled shell is even, and
$n_{W}=(n+1)(n+3)/2$ if it is odd, and these numbers define the heights of
steps in both the WCI and SCI regimes.

\begin{figure}[ht]
\centerline{\psfig{figure=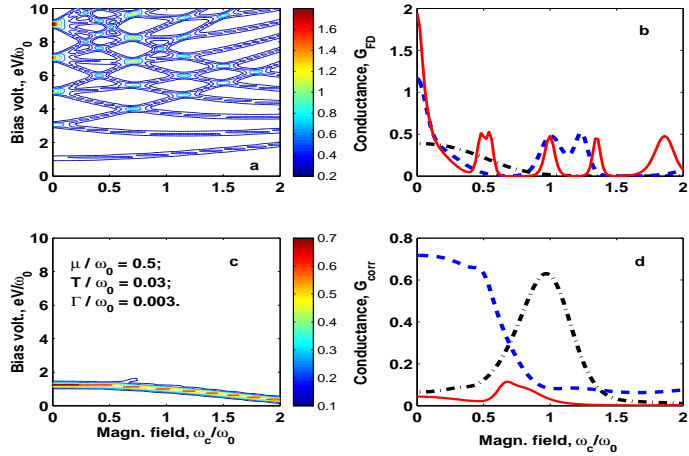,width=3.6in,height=2.4in,clip=}}
\caption{(Color online) The magnetoconductance $G=dJ/(eV)$ at the
WCI (left) and the SCI regimes (right). a)The contour plot for the
WCI conductance  displays the Fock-Darwin spectrum of the
parabolic QD in the magnetic field. b)The conductance in the WCI
regime, at fixed values of the bias voltage. The results for
$eV/\omega_0 = 1, 5, 9$ are presented by dot-dashed, dashed and
solid lines, correspondingly. c)The contour plot for the
self-consistent conductance at the SCI regime. d)The
self-consistent conductance at the SCI regime, at fixed values of
the bias voltage. The dot-dashed, dashed  and solid  lines present
the results for $eV/\omega_0 = 1, 1.25, 1.45$, correspondingly.
The parameters of the calculations are displayed in the panel
"c".} \label{6thFig}
\end{figure}

These features of the spectrum  can be traced by means of the
conductance measurements (see Fig.\ref{6thFig}). In particular,
the results for the differential conductance $dJ/d(eV)$ at the WCI
regime (Fig.\ref{6thFig}a) resembles very much the experimental
conductance discussed in Refs.\onlinecite{ota}. The increase of
the bias voltage allows to detect the degeneracy of quantum
transitions involved in the transport by  means of the
oscillations in the conductance (see Fig.\ref{6thFig}b).  The
large conductance magnitude at zero magnetic field corresponds to
the large number of the levels involved in the transport. With the
increase of the magnetic field at fixed bias voltage (solid line),
the conductance oscillates due to appearance and disappearance of
the shells. As was mentioned above, the increase of the magnetic
field brings into the CW states with a large magnetic quantum
number $m=n_{-}-n_{+}$ and pushes up the ones with $-m$. This
gives oscillations in the current (see Fig.\ref{2dFig}b), which
are suppressed, however, together with the current itself in the
SCI regime (see Fig.\ref{6thFig}c).

Let us discuss once more the physical origin of the current and
the conductance suppression by the correlations. 
An increase of the bias-voltage window involves
new and new levels (or, more accurately, transition energies) into
transportation of electrons through QD. In SCI regime (contrary to the WCI one)
all processes are developing under some constraints. First one
is that QD cannot contain more than one electron due to  a large energy 
gap between single- and two particle states (our model consideration 
 is performed in the limit $U \rightarrow \infty $).
 Second constraint comes from the normalization
condition: each channel participates in the transport with the
spectral weight $P_{\gamma}=N_0+N_{\gamma}< 1$, whereas the
population numbers  fulfil the sum rule $=N_0 + \sum_{\gamma
}N_{\gamma } = 1 $. This automatically leads to the conclusion that 
the more levels
 are involved, the less contribution each channel makes into the
current  and, correspondingly, to the conductance (see Fig.\ref{5thFig}). 
Furthermore, since
the parabolic QD has a spectrum with the shell structure, the number of
involved levels grows very fast. This explains why a strong
suppression of the transport by the correlations occurs in very narrow interval of
the bias voltage (see Fig.\ref{6thFig}d). Note that in spite of
 the strong suppression of the SCI conductance by the intra-dot
correlations, it is possible to observe a fine structure produced
by the shell structure as well at a certain values of the bias
voltage (see Fig.\ref{6thFig}d).

\section{Discussion and Summary}

In this work we perform an analysis of the non-linear transport
through QD, attached to two metallic contacts via
insulating/vacuum barriers,  in the regime of strong intra-dot
Coulomb interaction.  The calculation of the current and
conductance is based on the formula, derived by Meir and Wingreen
\cite{MW}. It expresses the current in terms of Green functions of
fermions in  a QD. 
 In the regime of strong Coulomb interaction 
the fermion variables are not good 
starting point for a constructing the perturbation theory with respect to 
the small transparency of the barriers. For this reason we use the approach where 
an active element (quantum dot) of
the device is supposed to be treated as good as possible (ideally,
exactly), whereas the coupling to the metallic contacts is treated
perturbatively. This leads to description in terms of {\em many}-electron operators; 
particulaly, here we find convenient to use the Hubbard operators.
We have re-written the
Meir-Wingreen's formula in terms of Green
functions (GFs), defined on Hubbard operators.
 The Wick's theorem does not work for such GFs. In
our approach this problem is solved by means of the extension of
the diagram technique for Hubbard-operator Green functions
(developed earlier for  a equilibrium in Ref.\onlinecite{ig}) to the
non-equilibrium states. In the first step, the exact functional equations for
the many-electron GFs are derived and used for the iterations in
the spirit of the Schwinger-Kadanoff-Baym approach (see also
Ref.\onlinecite{ig}). This gives rise to the diagrammatic
expansion. In the second step the analytical continuation of these equations to
the real-time axis transforms the system to the non-equilibrium
form.

  At the limit of infinitely  strong intra-dot Coulomb interaction,
the lowest approximation with respect to a small transparency of barriers
happen to be the mean field theory. In this theory the population
numbers of the intra-dot states and the energies of the transition between
these states have to be found self-consistently in non-equilibrium
(but stationary) states. Thus, the non-equilibrium mean field
theory for {\em multiple}-level QD has been constructed for non-zero
temperature (to the best of our knowledge this is the first attempt in
literature). It worth to note that a common practice is to
consider the population numbers of many-electron states in the
Gibbs' form (following, e.g., Ref.\onlinecite{MW}). In such  an approach the
population numbers do depend on the magnetic field, however the
dependence on the bias voltage is missing (cf
Refs.\onlinecite{datta1,datta2,asano}).

One of the questions we have addressed in the paper is to what
extent the levels (energies of the transitions) of the QD are
sensitive to the applied bias voltage $eV$.  We have found that this dependence 
is logarithmically weak (see Eq.\ref{log}). 
In contrast, the population numbers manifest much stronger and
highly non-trivial dependence on the bias voltage. At zero bias voltage they display the standard
Gibbs' statistics. However,  at the voltage higher than a certain
critical value (determined by the position of renormalized
transition energy) the population numbers start equalizing! This
occurs with only those of transitions which have energies within
the conducting window $(\mu_r, \mu_l )$. This result is obtained analytically
in simplified equations  and confirmed by the numerical solution
of  a complete system of non-linear integral equations.

The expression for the current for SCI regime that we have derived (Eq.(\ref{curr1})),
  demonstrates that the degree of the nonlinearity
of the transport is essentially determined by the spectral weights
$P_p$ of the dot states. Since they are
 just the sum of the population numbers, $P_p=N_0 + N_p$, 
 the non-trivial behavior of the
 latters  (see Eqs.(\ref{N_p}),(\ref{Phi}),(\ref{Np_and_N0})) 
is reflected in transport properties via
 a spectral weight of each conducting channel.

 Applying further these equations to the problem of the transport through a
two-dimensional parabolic quantum dot, we obtain a simple
expression which relates the height of the n-th step in the
current to the number of states participating in the transport. In
the WCI regime, at specific values of the magnetic filed,
Eq.(\ref{to}), we predict a drastic increase of the current
through the dot due to the shell effects.  Possibly, this effect 
can be used in applications (i.g., for 
magnetic-field sensitive switch on-off devices). 
  In  the SCI regime the transport is strongly suppressed. The origin 
of it again is connected with the spectral weights, regulating the contribution
of each conducting channel. First, the sum of population numbers 
is equal to one. Second, the number of levels within the conducting window 
grows very fast due to existence of the shell structure in the spectrum of
parabolic QD.
Third, in the limit $U \rightarrow \infty $ the QD cannot contain more than one electron. 
Therefore, the contribution of each channel unavoidably drops down when 
expanding by increasing voltage conducting window embraces
more and more transition energies.
 As a result, the current tends to  a certain limit, whereas the heights 
of the peaks in the conductance tend to zero. 
 Thus, we  conclude: while in the WCI regime the fine structures in the 
 conductance within one "diamond" can arise 
due to re-scattering/co-tunneling processes, in the SCI regime it has the other origin.
Namely, the role of single-electron levels is played by the energies of the transition between 
many-electron states, and the contribution of each transition 
into transport is regulated by the strength   
of its coupling to the conduction-electron subsystem and the spectral weights.

\section*{Acknowledgments}
This work was partly supported by Grant FIS2005-02796(MEC). I.S.
 thanks  UIB for the support and hospitality and R. G. N. is grateful to the
Ram\'{o}n y Cajal programme (Spain).

\appendix
\def\theequation{\thesection.\arabic{equation}}
\section{Relations between Fermi and Hubbard operators}
\setcounter{equation}{0}
\label{app1}

It is convenient to  use different notations for Hubbard operators
that describe Bose-like transitions (without change of particle
numbers) and  Fermi-like ones (between the states, differing by
one electron). The former and the later ones are denoted as ($Z$)
and ($X$), respectively:
\begin{eqnarray}
&&Z^{00} = \prod_{\gamma_{1}}\left(  1-d_{\gamma_{1}}^{\dagger}d_{\gamma_{1}}\right)=
\left|  0\right\rangle \left\langle 0\right|  ,\;\;
\label{vac}\\
&&Z^{\gamma\gamma} = \prod_{\gamma_{1}\neq \gamma}
\left(  1-d_{\gamma_{1}}^{\dagger}\gamma_{\gamma_{1}}\right)
d_{\gamma}^{\dagger}d_{\gamma}=\left|  \gamma\right\rangle \left\langle \gamma\right|,\;\;
\label{df_Z}\\
&&X^{0\gamma}=\prod_{\gamma_{1}\neq \gamma}
\left(  1-d_{\gamma_{1}}^{\dagger}d_{\gamma_{1}}\right)
d_{\gamma}=\left|  0\right\rangle \left\langle \gamma\right| ,
\label{df_X}\\
&&Z^{00}\left|0\right\rangle =\left|  0\right\rangle,\quad
Z^{\gamma\gamma}\left|  \gamma\right\rangle =\left|  \gamma\right\rangle ,
\label{act}
\end{eqnarray}
where Eq.(\ref{vac}) defines the vacuum state and the product is
taken over all single-particle states $\gamma_{1}$ of the QD
confining potential.
In this case the expansion of unity acquires
the form:
\begin{equation}
\hat{1}=Z^{00}+\sum_{\gamma}Z^{\gamma\gamma}.
\label{UnityExpansion}
\end{equation}
The expectation values $N_{0}=\left\langle Z^{00}\right\rangle $ and
$N_{\gamma}=\left\langle Z^{\gamma\gamma}\right\rangle $ with respect to state/ensemble of
interest are population numbers of corresponding many-electron states.
The Hamiltonians of closed QD in these terms become:%
\begin{equation}
H_{QD}^{WCI}=\sum_{\gamma}\varepsilon_{\gamma}d_{\gamma}^{\dagger}d_{\gamma},\;\;
H_{QD}^{SCI}=\varepsilon_{0}Z^{00}+\sum_{\gamma}\varepsilon_{\gamma}Z^{\gamma\gamma}.
\label{fp2}
\end{equation}
The WCI regime corresponds to $U\rightarrow0$, while the SCI regime takes place
at $U\rightarrow\infty$.
The energy of empty QD $\varepsilon_{0}$ is determined by the
potential depth and can be chosen arbitrarily.
While the eigenvalues $\varepsilon_{\gamma}$ are defined by shape of the potential,
the transition energies are
$\Delta_{\gamma}^{(0)}\equiv$ $\varepsilon_{\gamma}-\varepsilon_{0}$.
We recall that the position of the single-electron transitions
with respect to the electrochemical potentials is
essential for the transport properties of the system.

To write $H_{t}$ in these variables, we have to express the
annihilation (creation) operator $d_{\gamma}(d_{\gamma}^{\dagger})$ in terms of
Hubbard operators:
\begin{equation}
d_{\gamma}=\hat{1}\cdot d_{\gamma}\cdot\hat{1}=\left\langle 0\right|  d_{\gamma}\left|
\gamma\right\rangle X^{0\gamma}. \label{d}%
\end{equation}
\ Of course, the expansion (\ref{d}) should contain the transitions from one-
to two-particle, from two- to three-particle and so on configuration
${\Lambda^n}$
\begin{equation}
d_{\gamma}=\sum_{\left\{  \Lambda,n\right\}  }\left(  \Lambda^{(n)},n\left|
d_{\gamma}\right|  \Lambda^{(n+1)},n+1\right)  X^{\Lambda^{(n)}\Lambda^{(n+1)}},
\label{Xgen}%
\end{equation}
that defines the anticommutator relation
\begin{equation}
\hat{1}=\left\{  d_{\gamma},d_{\gamma}^{\dagger}\right\}  =\sum_{\left\{  \Lambda
,n\right\}  }Z^{\Lambda^{(n)}\Lambda^{(n)}}. \label{fullExpUnity}%
\end{equation}
However, in the limit $U\rightarrow \infty$
the above equation is reduced to Eq.(\ref{UnityExpansion}).
In other words, we consider the transport within the first
\ conducting ''diamond'' only,
in coordinates of gate and bias voltages. \ The $H_{t}$ remains unchanged
in the WCI regime, whereas in the SCI regime it can be written
in the form
\begin{equation}
H_{t}^{SCI}=\sum_{k,\sigma \in \lambda,\lambda,\gamma}
\left( v_{k\sigma,\gamma}^{\lambda}c_{\lambda k\sigma }^{\dagger }X^{0\gamma}+h.c.\right)
\end{equation}
where
\begin{eqnarray}
&v_{k\sigma;\gamma}^{\lambda}=\tilde{v}_{k\sigma
;\gamma}^{\lambda}
\langle 0|d_{\gamma}|\gamma\rangle,\nonumber \\
&v_{\gamma;k\sigma}^{\lambda}=\tilde{v}_{\gamma
;k\sigma}^{\lambda} \langle \gamma|d_{\gamma}^{\dagger}|0\rangle
\label{tmel}
\end{eqnarray}
The matrix element $v_{k\sigma;0\gamma}^{\lambda}$ contains
information on specific features of attachment between a quantum dot
and contacts. Here, $\lambda=l,r$ denotes the left and the right lead, respectively.
One may notice that Hubbard operators take into account
corresponding kinematic restrictions placed by the strong Coulomb interaction.
The prize for this is, however, non-trivial commutation relations:
\begin{eqnarray}
&\left\{  X^{0\gamma},X^{\gamma^{\prime}0}\right\} =
\delta_{\gamma \gamma^{\prime}}Z^{00}+Z^{\gamma^{\prime}\gamma},\;
[Z^{00},X^{0\gamma}]=X^{0\gamma},\nonumber\\
&[Z^{\gamma^{\prime}\gamma^{\prime\prime}},X^{0\gamma}]=
-\delta_{\gamma \gamma^{\prime}}X^{0\gamma^{\prime\prime}}.
\label{Commut}
\end{eqnarray}
Thus, our model Hamiltonian takes the form
\begin{eqnarray}
& H = \sum_{k\sigma \in \lambda,\lambda}\varepsilon_{\lambda k\sigma}
c_{\lambda k\sigma}^{\dagger}c_{\lambda k\sigma}+\varepsilon_0Z^{00}+
\sum_{\gamma}\varepsilon_{\gamma}Z^{\gamma\gamma}\nonumber\\
& + \sum_{k\sigma \in \lambda,\lambda,\gamma}
\left( v_{k\sigma,\gamma}^{\lambda}c_{\lambda k\sigma}^{\dagger }X^{0\gamma}+h.c.\right)
\label{fp7a}
\end{eqnarray}
We assume also that
$\{c_{\lambda k\sigma},c_{\lambda^{\prime}k^{\prime }\alpha^{\prime}}^{\dagger}\}=
\delta_{k,k^{\prime}}\delta_{\sigma,\sigma^{\prime}}
\delta_{\lambda,\lambda^{\prime}},\,\,
\{c_{\lambda k\sigma},d_{\gamma}^{\dagger}\}=0$.

As it was discussed in the Introduction, the particular sector of
the transitions ($0\Longleftrightarrow 1$) is remarkable due to
the fact that the bare energies $\varepsilon_{\gamma}$ and the
tunnelling matrix elements coincide in both the WCI and the SCI
regimes. This opens a nice opportunity to study the role played by
the strong Coulomb interaction in formation of transport
properties in a most refined form. Indeed, in the SCI case the
electrons in QD can "see" the conduction electrons in the attached
contacts only through the prism of the in-dot kinematic
constraints. This results in the renormalization of the transition
energies.

\section{Current}
\setcounter{equation}{0}
\label{app2}

Here we re-derive the expression for the current
obtained by Meir and Wingreen \cite{MW} in terms of
Green functions (GFs) for the Hubbard operators. Thus,
\begin{eqnarray}
&&J_l=\frac{ie}{\hbar}\sum_{k\sigma \in l,\gamma} \left\{
\tilde{v}_{k\sigma,0\gamma}^l\left\langle
c_{lk\sigma}^{\dagger}d_{\gamma}\right\rangle -\tilde{v}_{\gamma
0,k\sigma}^l\left\langle
d_{\gamma}^{\dagger}c_{lk\sigma}\right\rangle
\right\}  \nonumber\\
&&  =\frac{ie}{\hbar}\sum_{k\sigma \in l,\gamma}
\left\{  v_{k\sigma,\gamma}^l\left\langle c_{lk\sigma}^{\dagger}X^{0\gamma}\right\rangle
-v_{\gamma,k\sigma}^l
\left\langle X^{\gamma 0}c_{lk\sigma}\right\rangle \right\}  \nonumber\\
&&  =\frac{e}{\hbar}\sum_{k\sigma \in l,\gamma}
\int d\omega\left\{v_{k\sigma,\gamma}^lG_{\gamma,k\sigma}^{<}(\omega)
-v_{\gamma,k\sigma}^lG_{k\sigma,\gamma}^{<}(\omega)\right\}\nonumber\\.
\label{c1b}
\end{eqnarray}
where the matrix elements
$v_{k\sigma,\gamma}^l (v_{\gamma,k\sigma}^l$) are defined by Eq.(\ref{tmel}).
Equations for the Green functions
$G_{\gamma,k\sigma}$ and $G_{k\sigma,\gamma}$
on the Keldysh contour are
\begin{eqnarray}
G_{\gamma,k\sigma}(\tau,\tau^{\prime}) & =
\int_{C}d\tau_{1}G_{\gamma, \gamma^{\prime}}(\tau,\tau_{1})v_{\gamma^{\prime},k\sigma}^l
C_{k\sigma}^{(0)}(\tau_{1},\tau^{\prime})
\label{c2a}\\
G_{k\sigma,\gamma}(\tau,\tau^{\prime}) & =
\int_{C}d\tau_{1}C_{k\sigma}^{(0)}(\tau,\tau_{1})v_{k\sigma,\gamma^{\prime}}^l
G_{\gamma^{\prime},\gamma}(\tau,\tau_{1})
\label{c2b}%
\end{eqnarray}
Here $C_{k\sigma}^{(0)}(\tau_{1},\tau^{\prime})$ is a bare GF of
conduction electrons in the left lead.
Expressing GFs $G$ via integrals on real-time
axes and making a Fourier transformation with respect to a difference of times
(the steady state is considered only!), we obtain:
\begin{eqnarray}
G_{\gamma,k\sigma}^{<}(\omega)& = &G_{\gamma, \gamma^{\prime}}^{<}(\omega)
v_{\gamma^{\prime},k\sigma}^lC_{k\sigma}^{(0)A}(\omega)+\nonumber\\
& + &G_{\gamma, \gamma^{\prime}}^{R}(\omega)v_{\gamma^{\prime},k\sigma}^lC_{k\sigma}^{(0)<}(\omega)
\nonumber\\
G_{k\sigma,\gamma}^{<}(\omega)&=&C_{k\sigma}^{(0)<}(\omega)
v_{k\sigma,\gamma^{\prime}}^lG_{\gamma^{\prime},\gamma}^{A}(\omega)+\nonumber\\
& + &C_{k\sigma}^{(0)R}(\omega)v_{k\sigma,\gamma^{\prime}}^l
G_{\gamma^{\prime},\gamma}^{<}(\omega)\nonumber\\
\label{c5b}
\end{eqnarray}
Here, the Fourier transformation is defined as:
\begin{equation}
G(t)=\int\frac{d\omega}{2\pi}e^{-i\omega t}G(\omega),G(\omega)=\int
dte^{i\omega t}G(t).\label{Fourier}%
\end{equation}

Inserting Eqs.(\ref{c5b}) into Eq.(\ref{c1b}), we
have:
\begin{eqnarray}
J_{l} &  =\frac{e}{\hbar}\int d\omega\left[ G_{\gamma, \gamma^{\prime}}^{<}
(\omega)V_{\gamma^{\prime},\gamma}^{l,A}(\omega)+G_{\gamma, \gamma^{\prime}}^{R}(\omega)
V_{\gamma^{\prime}\gamma}^{l,<}(\omega)\right]  \nonumber\\
&  -\frac{e}{\hbar}\int d\omega\left[ V_{\gamma \gamma^{\prime}}^{l,<}(\omega)
G_{\gamma, \gamma^{\prime}}^{A}(\omega)+V_{\gamma \gamma^{\prime}}^{l,R}(\omega)
G_{\gamma, \gamma^{\prime}}^{<}(\omega)\right]
\label{curr2}
\end{eqnarray}
Here, we have introduced the effective
interaction with conduction electrons in the lead $\lambda=l$:
\begin{equation}
V_{\gamma^{\prime}\gamma}^{\lambda,\eta}(\omega)=\sum_{k\sigma \in \lambda}
v_{\gamma^{\prime},k\sigma}^{\lambda}C_{k\sigma}^{(0)\eta}(\omega)
v_{k\sigma,\gamma}^{\lambda},
\label{effin}
\end{equation}
where $\eta \equiv \{R,A,>,<\}$.
Since in the steady state the current is homogeneous, i.e.,
$J_{l}=(J_{l}+J_{l})/2=$ $(J_{l}-J_{r})/2$, we can symmetrize Eq.(\ref{curr2}):
\begin{eqnarray}
J_{l}=&&\frac{e}{2\hbar}\int d\omega
[(V_{\gamma^{\prime},\gamma}^{l,A}(\omega)
-V_{\gamma^{\prime},\gamma}^{l,R}(\omega))-\nonumber\\
&& -( V_{\gamma^{\prime},\gamma}^{r,A}(\omega)-V_{\gamma^{\prime},\gamma}^{r,R}(\omega))]
G_{\gamma \gamma^{\prime}}^{<}(\omega) +\\
&&+\frac{e}{2\hbar}\int d\omega(V_{\gamma^{\prime},\gamma}^{l,<}(\omega)
-V_{\gamma^{\prime},\gamma}^{r,<}(\omega))
\left(  G_{\gamma \gamma^{\prime}}^{R}(\omega)-
G_{\gamma \gamma^{\prime}}^{A}(\omega)\right)\nonumber
\label{curr3}
\end{eqnarray}

Let us calculate the effective interaction $V_{\gamma^\prime,\gamma}^{l,\eta}(\omega).$
The conduction-electron ''lesser'' and ''greater'' GFs for the left lead
($\lambda=l$)
or for the right one ($\lambda=r$) are:%
\begin{eqnarray*}
C_{k\sigma}^{(0)<}(t,t^{\prime}) &\equiv& i\left\langle
c_{\lambda k\sigma}^{\dagger}(t^{\prime})c_{\lambda k\sigma}(t)\right\rangle^{(0)}=
ie^{-i\varepsilon_{\lambda k\sigma}(t-t^{\prime})}
f_{\lambda}(\varepsilon_{\lambda k\sigma})\\
C_{k\sigma}^{(0)>}(t,t^{\prime}) & \equiv &-
i\left\langle c_{\lambda k\sigma}(t)c_{\lambda k\sigma}^{\dagger}(t^{\prime})
\right\rangle^{(0)}=\nonumber\\
& = & -ie^{-i\varepsilon_{\lambda k\sigma}(t-t^{\prime})}
f_{\lambda}(-\varepsilon_{\lambda k\sigma})\\
C_{k\sigma}^{(0)<}(\omega)& = &i\int d(t-t^{\prime})
e^{i\left(\omega-\varepsilon_{\lambda k\sigma}\right)  (t-t^{\prime})}
f_{\lambda}(\varepsilon_{\lambda k\sigma})=\nonumber\\
& = &2\pi i\delta\left( \omega-\varepsilon_{\lambda k\sigma}\right)
f_{\lambda}(\omega)\\
C_{k\sigma}^{(0)>}(\omega)& = &-i\int d(t-t^{\prime})
e^{i\left(\omega - \varepsilon_{\lambda k\sigma}\right)(t-t^{\prime})}
f_{\lambda}(-\varepsilon_{\lambda k\sigma})=\nonumber\\
& = & -2\pi i\delta\left(\omega-\varepsilon_{\lambda k\sigma}\right)
f_{\lambda}(-\omega)
\end{eqnarray*}
For the retarded and advanced GFs we have
\begin{eqnarray*}
C_{k\sigma}^{(0)R}(t,t^{\prime})  & \equiv &
-i\theta(t-t^{\prime})\left\langle \left\{c_{\lambda k\sigma}(t),
c_{\lambda k\sigma}^{\dagger}(t^{\prime})\right\}  \right\rangle _{\lambda}^{(0)} = \\
& = & -i\theta(t-t^{\prime})e^{-i\varepsilon_{\lambda k\sigma}(t-t^{\prime})}\\
C_{k\sigma}^{(0)R}(\omega) & = &-i\int d(t-t^{\prime})\theta(t-t^{\prime})
e^{i\left(\omega-\varepsilon_{\lambda k\sigma}\right)(t-t^{\prime})}= \\
& = &\frac{1}{\omega-\varepsilon_{\lambda k\sigma}+i\delta}\\
C_{k\sigma}^{(0)A}(t,t^{\prime})& \equiv & i\theta(-t+t^{\prime})
\left\langle \left\{c_{\lambda k\sigma}(t),c_{\lambda k\sigma}^{\dagger}
(t^{\prime})\right\}  \right\rangle _{\lambda}^{(0)}=\\
& = &i\theta(-t+t^{\prime})e^{-i\varepsilon_{\lambda k\sigma}(t-t^{\prime})};\\
C_{k\sigma}^{(0)A}(\omega)& = &i\int d(t-t^{\prime})\theta(-t+t^{\prime})
e^{i\left( \omega-\varepsilon_{\lambda k\sigma}\right)(t-t^{\prime})}=\\
& = &\frac{1}{\omega-\varepsilon_{\lambda k\sigma}-i\delta}.
\end{eqnarray*}
As a result, we have
\begin{eqnarray*}
&&\left(V_{\gamma^{\prime}\gamma}^{\lambda,A}(\omega)-
V_{\gamma^{\prime}\gamma}^{\lambda,R}(\omega)\right)=\\
& = &  \sum_{k\sigma \in \lambda}v_{\gamma^{\prime}, k\sigma}^{\lambda}
\left[C_{k\sigma}^{(0)A}(\omega)-C_{k\sigma}^{(0)R}(\omega)\right]
v_{k\sigma,\gamma}^{\lambda}=\\
& = & 2\pi i\sum_{k\sigma \in \lambda}v_{\gamma^{\prime},k\sigma}^{\lambda}
\delta(\omega-\varepsilon_{\lambda k\sigma})v_{k\sigma,\gamma}^{\lambda}=
2i\Gamma_{\gamma^{\prime}\gamma}^{\lambda}(\omega)
\end{eqnarray*}
and
\[
\left(V_{\gamma^{\prime}\gamma}^{l,<}(\omega)-V_{\gamma^{\prime}\gamma}^{r,<}(\omega)\right)
=2i\left(f_{l}(\omega)\Gamma_{\gamma^{\prime}\gamma}^{l}(\omega)-
f_{r}(\omega)\Gamma_{\gamma^{\prime}\gamma}^{r}(\omega)\right).
\]

With the aid of above equations we can rewrite Eq.(B7)
in terms of the width functions:
\begin{eqnarray}
&&J_{l}=\frac{ie}{\hbar}\sum_{\gamma \gamma^{\prime}}\int d\omega\left\{ \left[
\Gamma_{\gamma}^{\prime}\gamma^{l}(\omega)-\Gamma_{\gamma^{\prime}\gamma}^{r}(\omega)\right]
G_{\gamma \gamma^{\prime}}^{<}(\omega)+\right. \nonumber\\
&& + \left.  \left(  f_{l}(\omega)\Gamma_{\gamma^{\prime}\gamma}^{l}(\omega)-f_{r}
(\omega)\Gamma_{\gamma^{\prime}\gamma}^{r}(\omega)\right)
\left(  G_{\gamma \gamma^{\prime}}^{R}(\omega)-
G_{\gamma \gamma^{\prime}}^{A}(\omega)\right)  \right\}. \nonumber\\
\label{App1_curr}
\end{eqnarray}

\end{document}